\documentclass{aa}
\usepackage{graphicx}
\usepackage{txfonts}

\begin{document}

\title{Magnetic fields in Local Group dwarf irregulars$^*$}

\author{
K. T. Chy\.zy\inst{1}
\and M. We\.zgowiec\inst{1,3}
 \and R. Beck\inst{2}
 \and D. J. Bomans\inst{3}
}
\institute{Obserwatorium Astronomiczne Uniwersytetu
Jagiello\'nskiego, ul. Orla 171, 30-244 Krak\'ow, Poland
\and Max-Planck-Institut f\"ur Radioastronomie, Auf dem
H\"ugel 69, 53121 Bonn, Germany
\and Ruhr-Universit\"at Bochum, Universit\"atsstrasse 150, 44780 Bochum, Germany}
\offprints{K.T. Chy\.zy}
\mail{chris@oa.uj.edu.pl\\
$^*$Based on observations with the 100-m telescope at Effelsberg
operated by the Max-Planck-Institut f\"ur Radioastronomie (MPIfR) on behalf
of the Max-Planck-Gesellschaft.}
\date{Received date/ Accepted date}
\titlerunning{Magnetic fields in Local Group dwarfs}
\authorrunning{K.T. Chy\.zy et al.}

\abstract
{}
{
We wish to clarify whether strong magnetic fields can be effectively generated
in typically low-mass dwarf galaxies and to assess the role of dwarf galaxies
in the magnetization of the Universe.
}
{ We performed a search for radio emission and magnetic fields in an
unbiased sample of 12 Local Group (LG) irregular and dwarf irregular
galaxies with the 100-m Effelsberg telescope at 2.64\,GHz. Three
galaxies were detected. A higher frequency (4.85\,GHz) was used
to search for polarized emission in five dwarfs that are the most luminous
ones in the infrared domain, of which three were detected.}
{ Magnetic fields in LG dwarfs are weak, with a mean value of the
total field strength of $<4.2\pm 1.8\,\mu$G, three times lower than
in the normal spirals. The strongest field among all LG dwarfs of
$10\,\mu$G (at 2.64 GHz) is observed in the starburst dwarf IC\,10.
The production of total magnetic fields in dwarf systems appears to
be regulated mainly by the star-formation surface density (with the
power-law exponent of $0.30\pm0.04$) or by the gas surface density
(with the exponent $0.47\pm0.09$). In addition, we find
systematically stronger fields in objects of higher global
star-formation rate. The dwarf galaxies follow a similar
far-infrared relationship (with a slope of $0.91\pm0.08$) to that
determined for high surface brightness spiral galaxies. The
magnetic field strength in dwarf galaxies does not correlate with their 
maximum rotational velocity, indicating that a small-scale rather than a large-scale dynamo
process is responsible for producting magnetic fields in dwarfs.
If magnetization of the Universe by galactic outflows is coeval with
its metal enrichment, we show that more massive objects (such us Lyman break
galaxies) can efficiently magnetize the intergalactic medium with a
magnetic field strength of about 0.8\,nG out to a distance of
160--530\,kpc at redshifts 5--3, respectively. Magnetic fields that are 
several times weaker and shorter magnetization distances are expected for
primordial dwarf galaxies. We also predict that most star-forming
local dwarfs might have magnetized their surroundings up to a field strength 
about 0.1\,$\mu$G within about a 5\,kpc distance. }
{ Strong magnetic fields ($>6\,\mu$G) are observed only in dwarfs of
extreme characteristics (e.g. NGC\,4449, NGC\,1569, and the LG
dwarf IC\,10). They are all starbursts and more evolved
objects of statistically much higher metallicity and global
star-formation rate than the majority of the LG dwarf population.
Typical LG dwarfs are unsuitable objects for the efficient supply of
magnetic fields to the intergalactic medium. }

\maketitle

\section{Introduction}

Dwarf galaxies are the most numerous population of galaxies in the Universe
(Grebel~\cite{grebel01}) and according to the hierarchical clustering scenario,
they were the primary building blocks of more massive galaxies in the past.
They play a key role in one of the most puzzling problems in
the $\Lambda$CDM picture of galaxy formation, namely the ``missing
satellites problem'' (e.g. Kravtsov~\cite{kravtsov10}).
Much effort has been made to understand the origin of star formation bursts,
as well as the formation of filamentary structures in dwarfs (Hunter~\cite{hunter02},
Hunter \& Gallagher~\cite{hunter90}). In contrarast to the massive spiral galaxies, the star
formation activity in dwarfs occurs without any strong influence from density waves
(Hunter et al.~\cite{hunter98}) and usually develops stochastically.

Magnetic fields can bee especially important in these low-mass galaxies  
because of their lower gravitational potential and the greater
ability of their gas to escape via galactic winds. These fields enable
magnetic fields to be supplied to the intergalactic
medium (IGM) at early cosmological epochs (Kronberg et 
al.~\cite{kronberg99}, Bertone et al.~\cite{bertone06}). However, the generation of 
magnetic fields in dwarf galaxies by a classical
large-scale dynamo can be inefficient because of the slow or
chaotic rotation and hence low differential rotation of these galaxies
(Chy\.zy et al.~\cite{chyzy03}). Star formation activity would
produce magnetic fields by generating a small-scale dynamo (Zeldovich et 
al.~\cite{zeldovich90}), but the magnetic fields and star formation
would then be related in a nonlinear way (Chy\.zy~\cite{chyzy08}). Thus, it is
unclear whether strong magnetic fields could be effectively generated
in these low-mass galaxies and under stochastically generated star
formation the large-scale structures of magnetic fields could be developed.

Surprisingly, very strong magnetic fields were discovered in the optically
bright dwarf irregular galaxy NGC\,4449, with the total
field intensity of about $12\,\mu$G and a regular component of up to $8\,\mu$G
(Chy\.zy et al.~\cite{chyzy00}). Similar fields were detected in
NGC\,1569 (Kepley et al.~\cite{kepley10}). Weaker total fields, in the
range of $5-7\,\mu$G, were also observed in some other dwarfs as
NGC\,6822 (Chy\.zy et al.~\cite{chyzy03}), IC\,10 (Chy\.zy et 
al.~\cite{chyzy03}), and the Large Magellanic Cloud (LMC: Gaensler et al.~\cite{gaensler05}). 
In the Small Magellanic Cloud (SMC), a weak total field
of about $3\,\mu$G was found, partly on large-scales (Mao et
al.~\cite{mao08}). Since all these galaxies are optically bright and
nearby objects, the detection of magnetic fields in them may be influenced by strong
selection effects. The question then arises of whether they represent a typical 
sample of dwarf galaxies and the typical conditions for
dynamo processes to occur.

To date, there has been no systematic study of diffuse nonthermal
radio emission and polarization in a uniformly selected sample of dwarf
galaxies with diverse star formation activities, kinematics, masses, and gas contents.
One ideal target for such an investigation is the Local Group (LG), which contains a 
mixture of small irregular and dwarf galaxies around two giant spirals.
Many dwarfs in the Local Group are star-forming objects (Mateo~\cite{mateo98}) with
a wide range of star-forming activity (Hunter \& Elmegreen~\cite{hunter04},
Dolphin et al.~\cite{dolphin05}, Tolstoy et al.~\cite{tolstoy09}). As a
complete sample of local galaxies, LG dwarfs can be studied to provide reliable
statistical insight into association of magnetic fields with other galactic
properties. They also provide a unique opportunity to investigate the
relationship between magnetic fields in dwarfs and those in larger
stellar systems, giving valuable estimates of the efficiency of galactic dynamo
processes working in low-mass objects.

In this paper, we report the results of a systematic attempt to detect
diffuse radio emission and magnetic fields in LG irregular and dwarf irregular
galaxies. This sensitive study was conducted with the 100-m Effelsberg radio
telescope at 4.85~GHz and 2.64~GHz. In the next section, we describe the criteria
used to build our sample of dwarf galaxies and in Sect.~\ref{s:observations}
provide details of the observations and data reduction process. The radio maps,
derived radio emission fluxes, and estimates of magnetic field strengths are
presented in Sect.~\ref{s:results}. In Sect.~\ref{s:discussion} we 
investigate how magnetic fields might possibly be
influenced by the global and local star formation rate, galactic mass, rotation,
metallicity, and star formation history. We also investigate
the possibility of the Universe being magnetized by outflows from dwarfs
using our current knowledge on the IGM metal enrichment. We then 
discuss the prevalence of large-scale magnetic fields in dwarfs, and construct
a radio-infrared diagram to check whether dwarfs deviate from the general
correlation trend followed by optically bright spiral galaxies. We summarize our studies in
Sect.~\ref{s:summary}.

\section{The sample}
\label{s:sample}

The Local Group consists of about 41 members, the most massive and optically bright
of which are the three spiral systems the Milky Way, the M\,31, and the small M\,33.
Less significant members include seven galaxies of irregular type (Irr), e.g., LMC,
and 31 dwarfs. A subgroup of 14 dwarfs, called dwarf irregulars (dIrr), are low-mass
objects ($M\le 10^{9}\,M_{\sun}$) but like irregulars are gas-rich and display 
evidence of current or recent star formation (Mateo~\cite{mateo98}, Tolstoy et 
al.~\cite{tolstoy09}). The remaining 31 dwarfs, known as dwarf ellipticals (dE) and
dwarf spheroidals (dSph), are prime examples of gas-poor systems dominated by old
stellar populations, that show no signs of current star formation. They are low-luminosity
($M_V \ge -14$ mag), strongly dark-matter dominated systems, with total masses
of about $10^{7}\,M_{\sun}$ (Strigari et al.~\cite{strigari08}). Although dSph’s
experienced star formation over extended time intervals in their youth, today all
of them but one appear to be completely free of detectable interstellar material 
(Grebel et al.~\cite{grebel03}).
A new population of ultrafaint dSphs (UF dSphs) with absolute magnitude
$M_V \ge -6$ were observed by Martin et al. (\cite{martin08}). They are more metal-poor than
dSphs, their more luminous counterparts. Ursa Minor and Draco are examples of these systems,
whose stellar velocity dispersions are only a few kilometres per second, and
whose total masses of less than $10^{6}\,M_{\sun}$ are composed primarily of very old
stars.

For our systematic search for radio emission and magnetic fields, only
gas-rich systems that display evidence of star formation activity, hence irregular
and dwarf irregular galaxies, are suitable. Following Mateo~\cite{mateo98},
we do not distinguish between these two types of objects and call them hereafter
dwarf irregular galaxies.

There are in total 12 LG dwarf irregular galaxies of declination
larger than about $-25\degr$, and thus available for radio observations from the
site of the 100-m Effelsberg telescope. The details of the sample are presented in Table
\ref{t:objects}. To detect their radio emission and achieve an adequate
balance between resolution and sensitivity we used the 2.64\,GHz (11\,cm)
receiver of the Effelsberg telescope. Using a relatively low frequency, we also
minimized the contribution of radio thermal emission. At this frequency, the
telescope beam size of $4\farcm 6$ is still sufficient to probe the large-scale
radio emission of LG dwarfs, which are often of large angular size (up to 19\arcmin;
Table \ref{t:objects}). Radio interferometers that could provide data of higher resolution
are inappropriate for observations of these large and diffuse objects.

For a selected subgroup of 5 out of 12 dwarfs with a high star-formation rate (SFR)
(see Table \ref{t:bfields}), we also attempted to investigate in addition radio polarized
emission at 4.85\,GHz (6.2\,cm). We had observed three galaxies, NGC\,6822, IC\,10 (Chy\.zy 
et al.~\cite{chyzy03}), and IC\,1613 before with the Effelsberg telescope.
For the remaining two galaxies, Sextans\,A and Sextans\,B, we carried out separate
observations at 4.85\,GHz.

\section{Observations and data reduction}
\label{s:observations}

\begin{table*}[t]
\caption{Basic properties of the observed LG dwarfs$^*$}
\begin{center}
    \begin{tabular}{cccccccc}
\hline
\hline
Galaxy & Other & Type & \multicolumn{2}{c}{Optical position} & Apparent &  Linear & Distance \\
Name   &  Name &      & \hspace{5pt} $\textstyle\alpha_{2000}$ &
$\textstyle\delta_{2000}$ & size [\arcmin] & size [kpc] & [kpc]  \\
\hline
Aquarius & DDO 210 & dIrr & 20$^{\rm h}$46$^{\rm m}$51\fs8 \hspace{1pt} & $-12$\degr 50\arcmin 52\farcs 5 &
2.2$\times$1.1 & 0.61$\times$0.30 & 940 \\
GR 8 & DDO 155 & dIrr & 12$^{\rm h}$58$^{\rm m}$40\fs4 \hspace{1pt} & +14\degr 13\arcmin 02\farcs 9 &
1.1$\times$1 & 0.70$\times$0.64 & 2200 \\
IC 1613 & DDO 8 & Irr & 01$^{\rm h}$04$^{\rm m}$47\fs8 \hspace{1pt} & +02\degr 07\arcmin 04\farcs 0 &
16.2$\times$14.5 & 3.44$\times$3.08 & 730 \\
NGC 6822 & DDO 209 & Irr & 19$^{\rm h}$44$^{\rm m}$56\fs6 \hspace{1pt} & $-14$\degr 47\arcmin 21\farcs 4 &
15.5$\times$13.5 & 2.25$\times$1.96 & 500 \\
WLM & DDO 221 & Irr & 00$^{\rm h}$01$^{\rm m}$58\fs2 \hspace{1pt} & $-15$\degr 27\arcmin 39\farcs 3 &
11.5$\times$4 & 3.08$\times$1.07 & 920 \\
IC 10 & UGC 192 & Irr & 00$^{\rm h}$20$^{\rm m}$17\fs3 \hspace{1pt} & +59\degr 18\arcmin 13\farcs 6 &
6.8$\times$5.9 & 1.31$\times$1.13 & 660 \\
LGS 3 & PGC 3792 & dIrr/dSph & 01$^{\rm h}$03$^{\rm m}$55\fs0 \hspace{1pt} & +21\degr 53\arcmin 06\farcs 0 &
2$\times$2 & 0.36$\times$0.36 & 620 \\
SagDIG & PGC 63287 & dIrr & 19$^{\rm h}$30$^{\rm m}$00\fs0 \hspace{1pt} & $-17$\degr 40\arcmin 41\farcs 3 &
2.9$\times$2.1 & 0.88$\times$0.64 & 1040 \\
Sextans A & DDO 75 & dIrr & 10$^{\rm h}$11$^{\rm m}$00\fs8 \hspace{1pt} & $-04$\degr 41\arcmin 34\farcs 0 &
5.9$\times$4.9 & 2.26$\times$1.88 & 1320 \\
Sextans B & DDO 70 & dIrr & 10$^{\rm h}$00$^{\rm m}$00\fs1 \hspace{1pt} & +05\degr 19\arcmin 56\farcs 0 &
5.1$\times$3.5 & 2.02$\times$1.39 & 1360 \\
Leo A & DDO 69 & dIrr & 09$^{\rm h}$59$^{\rm m}$26\fs5 \hspace{1pt} & +30\degr 44\arcmin 47\farcs 0 &
5.1$\times$3.1 & 1.03$\times$0.62 & 690 \\
Pegasus & DDO216 & dIrr/dSph & 23$^{\rm h}$28$^{\rm m}$36\fs3 \hspace{1pt} & +14\degr 44\arcmin 34\farcs 5 &
5$\times$2.7 & 1.11$\times$0.60 & 760 \\
\hline
\end{tabular}
\label{t:objects}
\end{center}
$^*$ Data from Mateo~(\cite{mateo98}), Karachentsev~(\cite{karachentsev05}), and
the NED database.
\end{table*}

The Effelsberg observations were made in May 2007 (Table\,\ref{t:obs}). The
unbiased sample of 12 LG dwarfs were observed at 2.64\,GHz
using a single horn receiver. We scanned our objects alternatively along
the RA and Dec directions. The subgroup of two galaxies mentioned above were
additionally observed at 4.85\,GHz using a two-horn (with horn separation of
8\arcmin) system in the secondary focus of the radio telescope
(see Gioia et al.~\cite{gioia}). The coverages were obtained in this case in the
azimuth-elevation frame. At both frequencies, the horns were equipped with
two total-power receivers and an IF-polarimeter, resulting in four channels containing
the Stokes parameters I (two channels), Q, and U. The telescope pointing was corrected
by repeating cross-scans of a bright point source close to the observed galaxy at time
intervals of about 1.5\,h. The flux density scale was established by mapping the calibration
sources 3C\,138 and 3C\,286. Table~\ref{t:obs} presents the details of our
observations at 2.64\,GHz.

\begin{table*}[t]
\caption{Parameters of the radio observations of all galaxies studied at 2.64\,GHz.}
\begin{center}
\begin{tabular}{ccccc}
\hline
\hline
Galaxy & Map size  & r.m.s. in final map & No. if coverages & Total flux \\
       & arcmin$\times$arcmin & mJy/b.a. &                  & mJy \\
\hline
Aquarius & 30$\times$30 & 0.77 & 8 & $\le 1.2$ \\
GR 8 & 30$\times$30 & 0.54 & 15 & $\le 0.8$ \\
IC 1613 & 44$\times$44 & 1.10 & 11 & $17\pm2$ \\
NGC 6822 & 44$\times$44 & 1.20 & 8 & $120\pm20$ \\
WLM & 40$\times$40 & 1.40 & 11 & $\le 4.2$ \\
IC 10 & 36$\times$36 & 3.90 & 10 & $250\pm20$ \\
LGS 3 & 30$\times$30 & 0.49 & 10 & $\le 0.74$ \\
SagDIG & 30$\times$30 & 0.72 & 10 & $\le 1.1$ \\
Sextans A & 40$\times$40 & 1.70 & 18 & $\le 2.6$ \\
Sextans B & 40$\times$40 & 1.10 & 12 & $\le 1.7$ \\
Leo A & 40$\times$40 & 0.88 & 8 & $\le 1.3$ \\
Pegasus & 40$\times$40 & 0.81 & 12 & $\le 1.2$ \\
\hline
\end{tabular}
\end{center}
\label{t:obs}
\end{table*}

The data reduction was accomplished using the NOD2 data reduction package
(Haslam~\cite{haslam}). At 2.64~GHz, the obtained coverages in I, Q, and U
channels (from a single horn system) were combined using the spatial-frequency
weighting method (Emerson \& Gr\"ave~\cite{emerson}), followed by a digital
filtering process that removed the spatial frequencies corresponding to noisy
structures smaller than the telescope beam.

At 4.85~GHz (dual system used), we combined the data from the two horns,
using the `software beam switching' technique (Morsi \& Reich~\cite{morsi}), followed
by restoration of total intensity channel I (Emerson et al.~\cite{emerson2}).
We then combined I, Q, and U maps using the same procedure as for 2.64\,GHz and obtained the final maps
of total power, polarized intensity, polarization degree, and polarization position
angles.

\section{Results}
\label{s:results}

\subsection{Radio detections}
\label{s:detection}

Among the 12 observed LG dwarfs, we clearly detected extended radio
emission at 2.64\,GHz in three cases: IC\,10, NGC\,6822, and
IC\,1613. To estimate their radio fluxes reliably or their upper
limits in cases of no detection the emission from confusing
background sources had to be removed. For identification of
background sources, we compared our radio maps with those of 
the Condon~(\cite{condon}), NVSS\footnote{NVSS: the NRAO VLA Sky Survey at 1.4
GHz, Condon et al.~(\cite{condon98})}, and FIRST\footnote{FIRST:
Faint Images of the Radio Sky at 20 cm, Becker et 
al.~(\cite{becker95})} surveys. We then applied the `subtraction'
method (Chy\.zy et al.~\cite{chyzy03}) and removed all confusing
sources from the maps. For clear detections of dwarfs their total
fluxes were obtained by integrating the signal in polygonal areas
encompassing all visible radio emission. If there was no detection,
we defined the upper limit of flux density as 1.5 times the rms
noise level of the map multiplied by the number of beams covering
the optical extent of a given galaxy. These estimates are presented
in Table.~\ref{t:obs}. All the detected galaxies are presented below
and in Figs. 1-5, which also show background sources. Polarized
emission was detected in NGC\,6822, IC\,10 (Chy\.zy et
al.~\cite{chyzy03}), and IC\,1613 (Figs. 4, 5) all at 4.85\,GHz. 
To show the structure of the magnetic field projected on the sky
plane, we used the apparent B-vectors defined as E-vectors rotated by
$90\degr$. Faraday rotation was expected to be small at 4.85\,GHz.

\begin{figure}
\begin{center}
\resizebox{8cm}{!}{\includegraphics{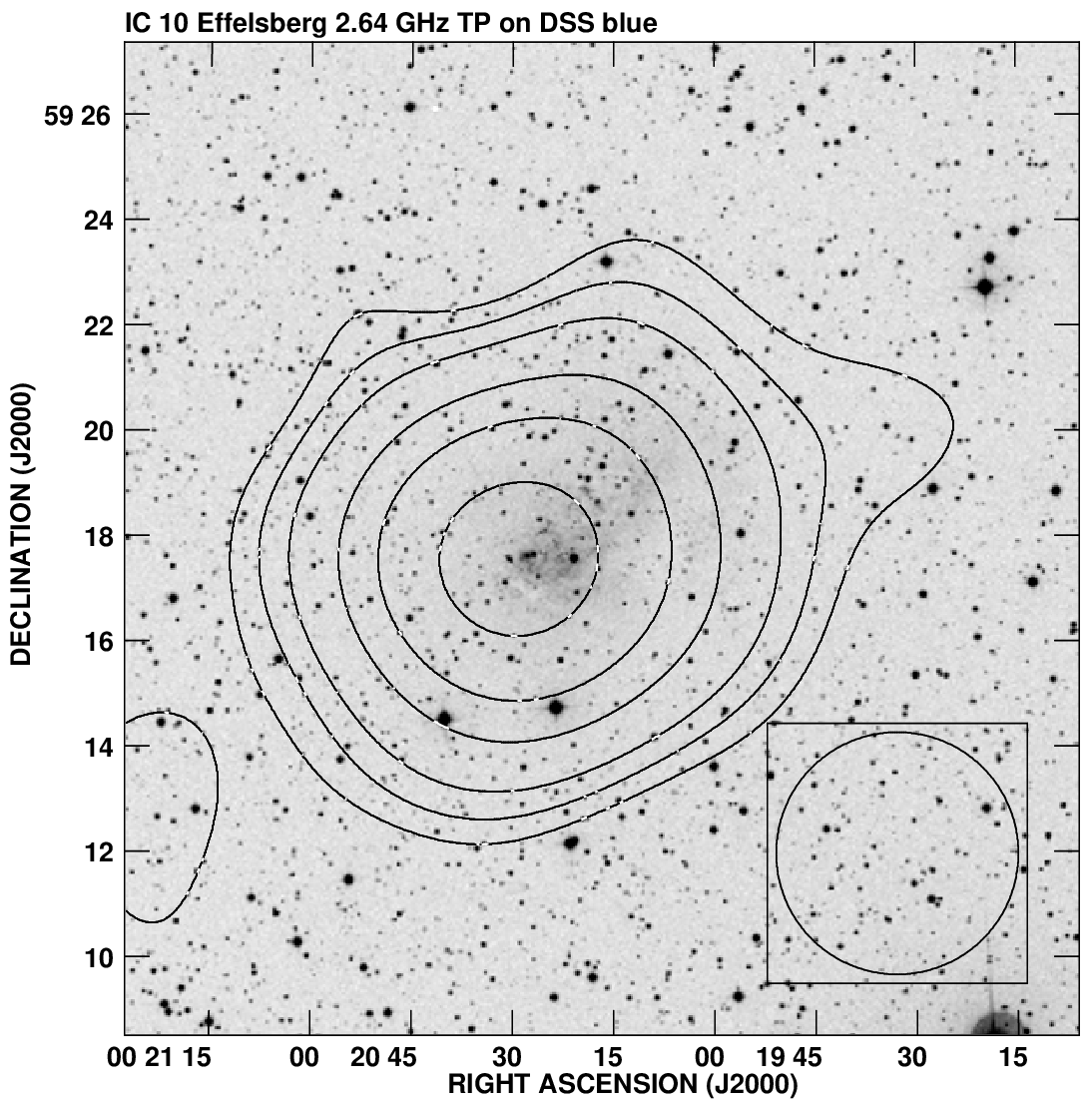}}
\end{center}
\caption{
The total power map of IC\,10 at 2.64~GHz overlaid onto the DSS
blue image. The contours are at 3, 5, 8, 16, 25, 40 $\times$ 3.9~mJy/b.a. The map
resolution is $4\farcm 6$. The beam size is shown in the bottom right corner of the figure.}
\label{ic10t}
\end{figure}

{\em IC\,10.} --- The map of total intensity at 2.64\,GHz
(Fig.~\ref{ic10t}) clearly detects, slightly resolved, and
elongated radio emission. It appears to be the strongest (250\,mJy)
source in our sample. The small extension to the west is caused by a
weak (9\,mJy) background source, also visible in the NVSS map at
1.4\,GHz. The emission is fully compatible with the 10.45\,GHz map
of Chy\.zy et al.~(\cite{chyzy03}). The spectral index between both
frequencies is about $0.35\pm0.05$, which is in good agreement with
Klein and Gr\"ave (\cite{klein86}). IC\,10
is experiencing a massive starburst, apparently triggered by
infalling \ion{H}{i} gas from the southeast (Grebel~\cite{grebel04}). 
The synchrotron emission detected in this portion
of the galaxy indicates gas compression (Chy\.zy et
al.~\cite{chyzy03}). According to several authors (see Grebel~\cite{grebel04} 
and references therein), the properties of this galaxy
suggest that it should be classified as a blue compact dwarf.

\begin{figure}
\begin{center}
\resizebox{8cm}{!}{\includegraphics{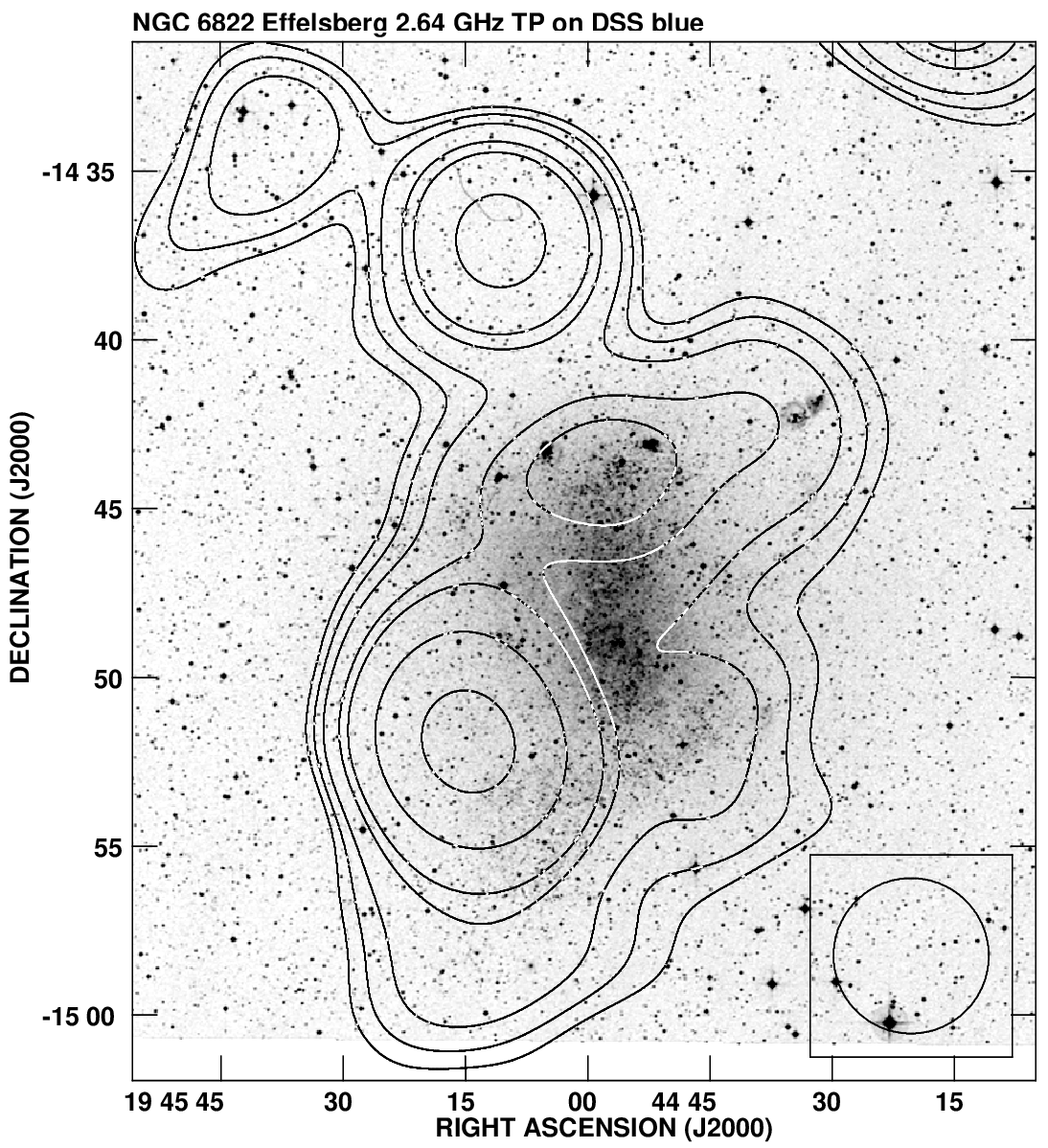}}
\end{center}
\caption{
The total power map of NGC\,6822 at 2.64~GHz overlaid onto the DSS
blue image. The contours are at 3, 5, 8, 15, 20, 40, 80 $\times$ 1.2~mJy/b.a. The map
resolution is $4\farcm 6$. The beam size is shown in the bottom right corner of the figure.}
\label{f:ngc6822}
\end{figure}

{\em NGC\,6822.} --- This galaxy has a three times lower total star
formation rate (based on H$\alpha$ observations) than IC\,10 (Woo et
al.~\cite{woo08}), but contains several well-defined supernovae
remnants and star-forming clumps (Chy\.zy et al.~\cite{chyzy03}; see
also Table \ref{t:bfields}). Our radio map of the total intensity of
NGC\,6822 at 2.64\,GHz (Fig.\,\ref{f:ngc6822}) shows significant
radio emission from this galaxy mainly associated with distinct
H$\alpha$ regions. The emission peaks at the positions RA$_{2000}$=$\rm
19^h45^m40^s$, Dec$_{2000}$=$-14\degr34\arcmin$, RA=$\rm
19^h45^m10^s$, Dec=$-14\degr37\arcmin$, and RA=$\rm 19^h45^m15^s$,
Dec=$-14\degr52\arcmin$ as well as the extension to the south are
due to background sources. The overall distribution of radio
emission corresponds well with our earlier observations at 4.85\,GHz
(Chy\.zy et al.~\cite{chyzy03}).

\begin{figure}
    \begin{center}
        \resizebox{8cm}{!}{\includegraphics{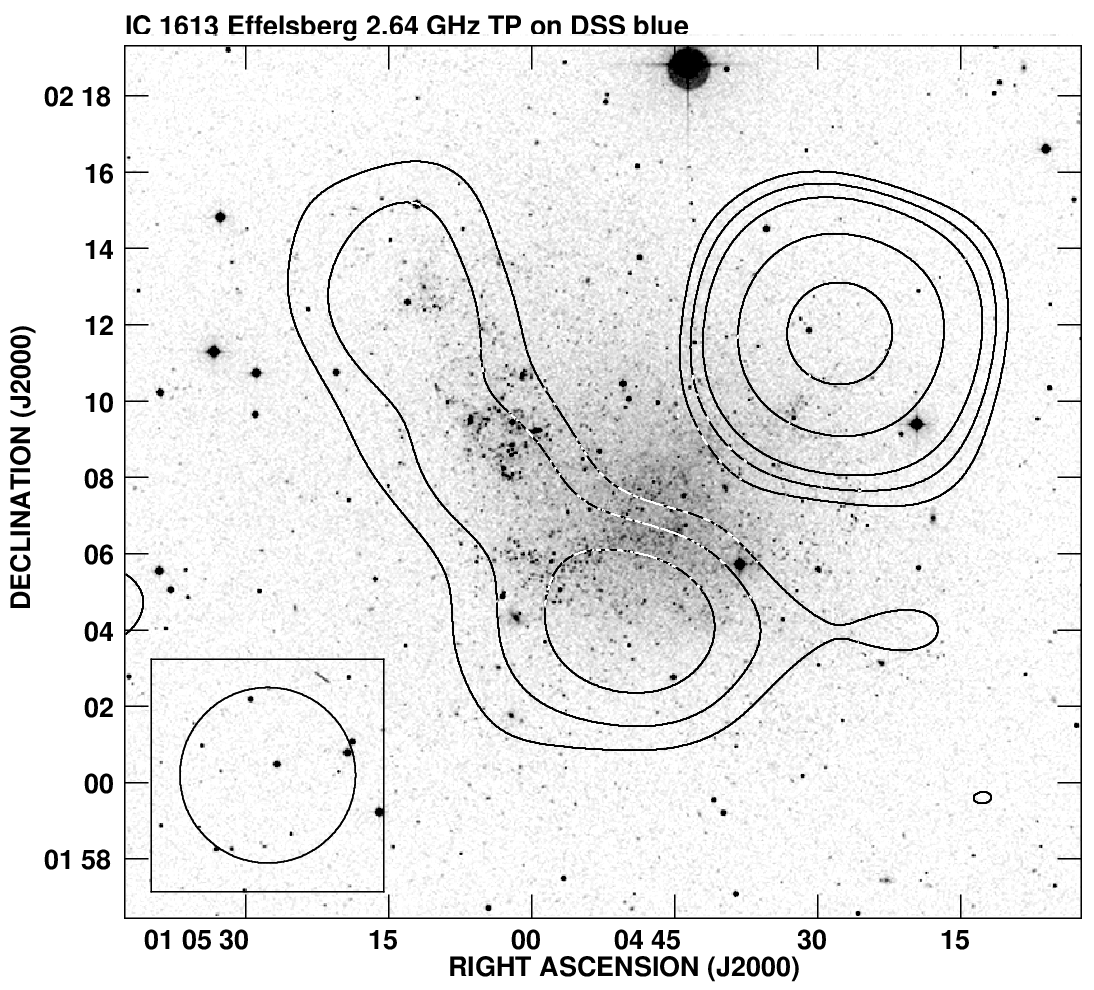}}
    \end{center}
    \caption{
    The total power map of IC\,1613 at 2.64\,GHz overlaid onto the DSS
    blue image. The contours are at 3, 5, 8, 20, 40 $\times$ 1.1~mJy/b.a. The map
    resolution is $4\farcm 6$. The beam size is shown in the bottom left corner of the figure.}
    \label{ic1613}
\end{figure}

\begin{figure}
\begin{center}
\resizebox{8cm}{!}{\includegraphics{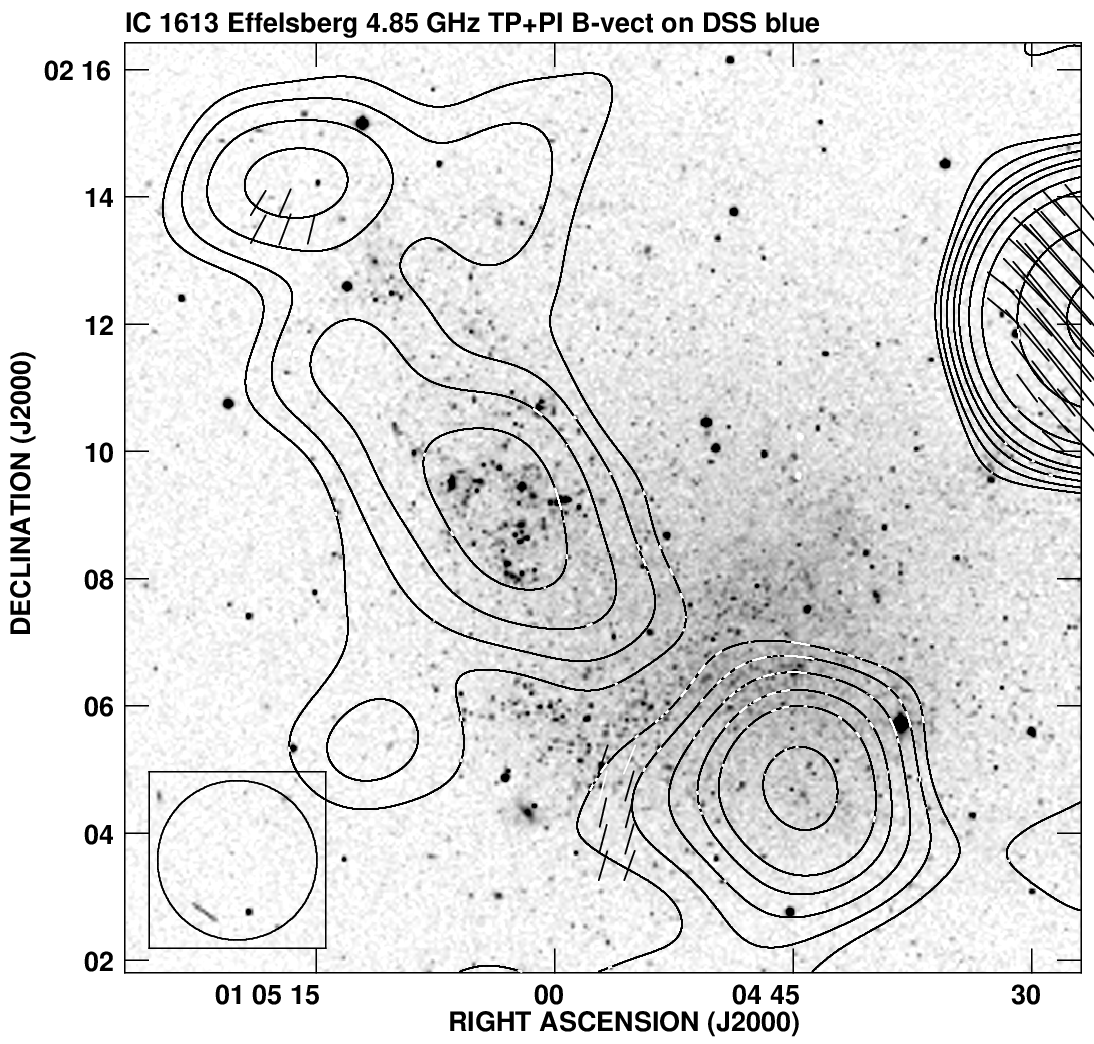}}
\end{center}
\caption{
The total power map of IC\,1613 at 4.85~GHz with apparent B-vectors of
polarized intensity overlaid onto the DSS blue image. The contours are at 3, 5,
8, 12, 16, 25, 40, 100, 200 $\times$ 0.3~mJy/b.a., and a vector of $1\arcmin$
length corresponds to the polarized intensity of 0.3~mJy/b.a. The map resolution
is $2\farcm 6$. The beam size is shown in the bottom left corner of the figure.}
\label{ic1613tp}
\end{figure}

\begin{figure}
    \begin{center}
        \resizebox{8cm}{!}{\includegraphics{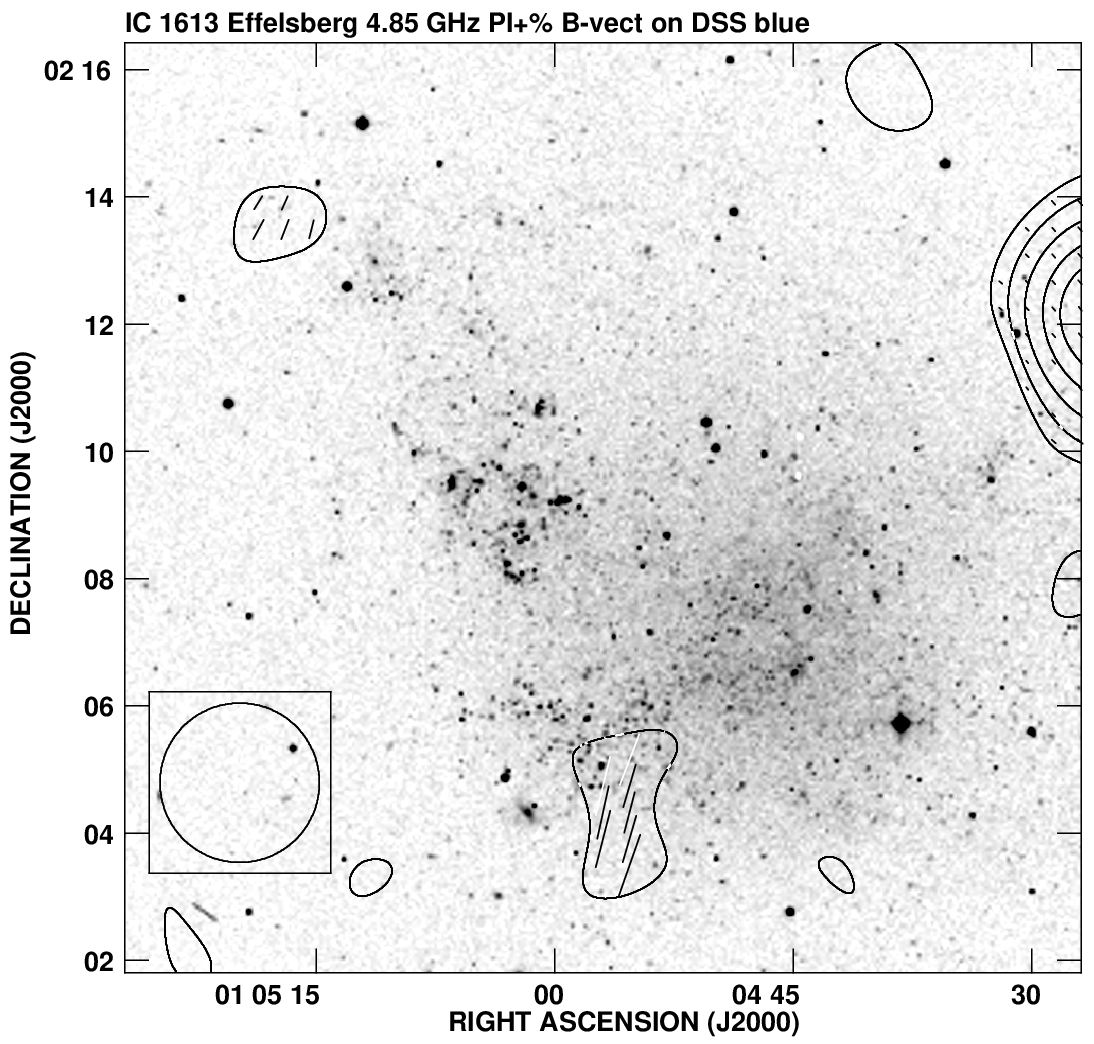}}
    \end{center}
\caption{
The map of polarized intensity of IC\,1613 at 4.85~GHz with apparent B-vectors
of polarization degree overlaid onto  the DSS blue image. The contours are at 3,
5, 8, 12, 16 $\times$ 0.05~mJy/b.a., and a vector of $1\arcmin$ length corresponds to
the polarization degree of 20\%. The map resolution is $2\farcm 6$. The beam size
is shown in the bottom left corner of the figure.}
\label{ic1613pi}
\end{figure}

{\em IC\,1613.} --- This galaxy seems to be a typical LSB Irr
evolving slowly in isolation without large bursts of star formation
during its entire lifetime (Skillman et al.~\cite{skillman03}). The
total power map at 2.64\,GHz (Fig.~\ref{ic1613}) reveals radio
emission mainly from areas associated with two distinct H$\alpha$
regions. The emission peaks at RA=$\rm 01^h04^m50^s$,
Dec=$02\degr04\arcmin$ and RA=$\rm 01^h04^m25^s$,
Dec=$02\degr12\arcmin$ are strong background sources. The higher
resolution map at 4.85\,GHz (Fig.~\ref{ic1613tp}) confirms these
findings, while revealing three more background sources (RA=$\rm
01^h05^m15^s$, Dec=$02\degr14\arcmin$ and RA=$\rm 01^h05^m12^s$,
Dec=$02\degr05\arcmin30\arcsec$ with an extension towards RA=$\rm
01^h04^m55^s$, Dec=$02\degr04\arcmin$ being a weak, though slightly
polarized background source). The main radio emission from
the galaxy comes from strong \ion{H}{ii} regions, whereas the rest
of the galaxy remains radio-quiet. The map of polarized intensity of
IC\,1613 (Fig.~\ref{ic1613pi}) shows only two faint patches of
emission in the northeastern and southern outskirts of the galaxy.
They are however most likely associated with the background sources
mentioned above.

\begin{figure}
\begin{center}
\resizebox{8cm}{!}{\includegraphics{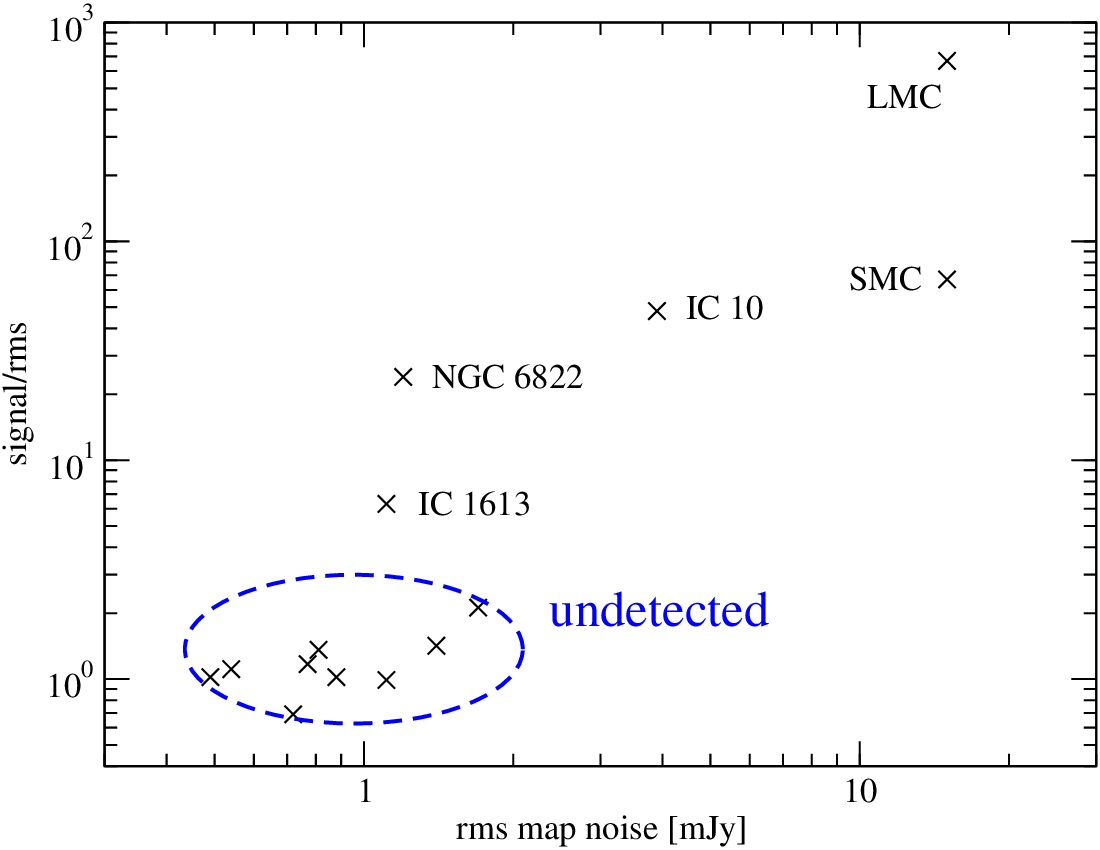}}
\end{center}
\caption{
Signal to r.m.s. noise level ratio versus the noise level  for our maps 
of LG dwarfs and for maps of the Magellanic Clouds (Haynes et al. 
\cite{haynes91})  at 2.64\,GHz. The undetected objects are marked.
}
\label{f:noise}
\end{figure}

To summarize, only 3 out of 12 sources (25\%) are detected at radio wavelengths.
The failed attempts to detect dwarfs are not due to the lower sensitivity level 
of their respective radio maps, as illustrated in Fig.\,\ref{f:noise}. They are
simply intrinsically weaker than the detected objects. This is not quite an
unexpected result as we analyse the volume {\em complete} sample of
LG dIrrs, unaffected by any selection bias. Although 75\% dwarfs
from our complete sample are undetected, they still provide
important information on the processes of magnetic field generation.
By comparing the undetected and detected LG dwarfs and by relating
them to other dwarfs observed so far (see Introduction), we can
statistically infer which properties of the galaxies influence the
radio emission and the generation of magnetic fields in dwarfs (see
Sect.\,\ref{s:discussion}).


\begin{table*}[ht]
\caption{Magnetic field estimates and physical parameters for our sample of LG
dwarfs and comparison dwarfs.}
\begin{center}
\begin{tabular}{crrrrrrr}
\hline
\hline
Galaxy name & B$_{tot}^{a}$  & SFR$^{b}$         & $\ion{H}{i}$ mass$^{c}$ & Total mass$^{d}$ &
S$_{60\mu m}$$^{e}$ & v$_{rot}$$^{f}$  & $\sigma_v^{g}$ \\
            & $\mu$G           & M$_{\sun}$ yr$^{-1}$ & 10$^{6}$ M$_{\sun}$   & 10$^{6}$ M$_{\sun}$ &
mJy                & km\,s$^{-1}$ & km\,s$^{-1}$ \\
\hline
Aquarius & $<4.5\pm$1.2 & 4.6$\times$10$^{-5}$ & 2.7 & 5.4 & 139 & 13 & 6.6\\
GR\,8     & $<3.6\pm$0.9 & 7.0$\times$10$^{-4}$ & 9.6 & 7.6 & 20 & 21  & 11.0 \\
IC\,1613 &  2.8$\pm$0.7 & 3.0$\times$10$^{-3}$ & 58  & 795 & 1420 & 37 & 8.5  \\
NGC\,6822 & 4.0$\pm$1.0 & 2.1$\times$10$^{-2}$ & 140 & 1640 & 47600 & 51 & 8.0\\
WLM      & $<3.9\pm$0.9 & 1.0$\times$10$^{-3}$ & 63  &  150 & 320 & 23 & 8.0 \\
IC\,10    &  9.7$\pm$2.0 & 6.0$\times$10$^{-2}$ & 98  & 1580 & 31200 & 47 & 8.0 \\
LGS\,3    & $<4.0\pm$1.0 & 2.5$\times$10$^{-6}$ & 0.2 &  13 & 75 & 18 & 9.0 \\
SagDIG   & $<4.1\pm$1.1 & 6.7$\times$10$^{-5}$ & 8.6 & 9.6 & 94 & 14 & 7.5 \\
Sextans\,A &$<3.1\pm$0.8 & 2.0$\times$10$^{-3}$ & 54  & 395 & 503 & 33 & 8.0 \\
Sextans\,B &$<2.8\pm$0.6 & 2.0$\times$10$^{-3}$ & 44  & 885 & 246 & 38 & 18.0 \\
Leo\,A     &$<4.4\pm$1.2 & 3.2$\times$10$^{-5}$ & 7.6 & 11 & 90 & 18 & 9.3 \\
Pegasus  &  $<3.7\pm$0.9 & 3.0$\times$10$^{-4}$ & 3.4 & 58 & 55 & 17 & 8.6 \\
\hline
LMC      & 4.3$\pm 1.0$ & 2.6$\times$10$^{-1}$ & 500 & 20000 & 8.29$\times$10$^{7}$ & 72 & 14.1 \\
SMC      & 3.2$\pm 1.0$ & 4.6$\times$10$^{-2}$ & 420 & 2400 & 7.45$\times$10$^{6}$ & 60 & 25.0 \\
NGC\,4449 & 9.3$\pm$2.0 & 4.7$\times$10$^{-1}$ & 2500&70000 & 36000 & 40 & 20.0 \\
NGC\,1569 & 14$\pm 3.0$ & 3.2$\times$10$^{-1}$ & 130 & 297 & 54400 & 42 & 21.3 \\
\hline
\end{tabular}
\label{t:bfields}
\end{center}
{\bf References.} $^{(a)}$  this paper and for LMC -- Gaensler et al.~(\cite{gaensler05}), SMC -- Mao
et al.~(\cite{mao08}), NGC\,4449 -- estimation based on 4.86\,GHz data (Chy\.zy et 
al.~\cite{chyzy00}), NGC\,1569 -- Kepley et al.~(\cite{kepley10});
$^{(b)}$ Woo et al.~(\cite{woo08}), if not available -- Hunter \& Elmegreen~(\cite{hunter04}),
for Aquarius and LGS\,3 estimated from infrared emission;
$^{(c)}$ Woo et al.~(\cite{woo08}); NGC\,1569 --  Stil \& Israel (\cite{stil02});
NGC\,4449  -- Hunter et al. (\cite{hunter99});
$^{(d)}$ Mateo~(\cite{mateo98});
$^{(e)}$IRAS (Helou \& Walker~\cite{iras}, Moshir et al.~\cite{iras2});
$^{(f)}$estimated rotational velocity defined as the maximum of rotational velocity and 2 times 
the galaxy velocity dispersion -- Woo et al.~(\cite{woo08}), for NGC\,4449  -- 
Valdez-Guti\'errez (\cite{valdez02});
$^{(g)}$velocity dispersion of ISM estimated from \ion{H}{i}: Mateo~(\cite{mateo98}); LMC -- Prevot et al. (\cite{prevot89}); 
SMC -- Staveley-Smith et al. (\cite{staveley97}); NGC\,1569 --  Stil
\& Israel (\cite{stil02});
for NGC\,4449  -- Hunter et al. (\cite{hunter99}).
\end{table*}

\subsection{Magnetic field strengths}
\label{s:magnetic}

After determining the radio emission flux or at least its upper
limit for all 12 LG dwarfs, we can calculate either the magnetic
field strength or its upper limit, respectively. We derive the thermal
contributions to the total radio fluxes mostly from
H$\alpha$ total fluxes. The classical model of \ion{H}{ii} regions
by Caplan \& Deharveng~(\cite{caplan86}) provides an estimate of the
radio thermal emission from H$\alpha$ fluxes, which we apply to the data 
at 2.64\,GHz. Extinction due to dust is low in these galaxies (Hunter 
\& Elmegreen~\cite{hunter04}), therefore H$\alpha$ clearly traces the bulk of
galactic star formation and can be used to estimate the SFRs from
simple linear scaling (i.e. Kennicutt~\cite{kennicutt98}). These SFRs
taken from the available literature are given in Table
\ref{t:bfields}. For Aquarius and LGS\,3, which are undetected in
H$\alpha$, the SFR and the expected radio thermal emission is
derived from the infrared emission (Kennicutt~\cite{kennicutt98}).
The thermal radio fluxes are subtracted from the total radio fluxes
(or their upper limits) to yield estimates of the synchrotron
emission.

To compute the equipartition magnetic field strengths, we employed
the formulae given by Beck \& Krause~(\cite{beckra}) (see Appendix 
for details). Since the presented calculations include the total 
magnetic field and its component perpendicular 
to the line of sight, as well as the pathlength through the medium, it is
required to take into account the geometries of the investigated galaxies and the 
contribution to the total field of both regular and random components.
As our aim is to search for the existence of magnetic fields and estimate 
their upper limits, it is enough to assume ellipsoidal geometries for the studied galaxies,
using their minor axes as the synchrotron pathlength. 
To estimate the uncertainties in the estimated field strengths, we
applied a variation in all assumed parameters of 50\%. 
Similar estimates of the magnetic field strengths were obtained at
4.85\,GHz for galaxies detected at this frequency. They give values
within the uncertainty range of the strengths derived from the
2.64\,GHz data.

For comparison, Table \ref{t:bfields} also presents data for the LMC
and SMC dIrrs, which are also LG members but are situated in the
southern sky (thus not included in our unbiased LG sample). We also
present data for NGC\,1569 and NGC\,4449, dIrrs not belonging to the
LG but with significant magnetic fields. NGC\,1569 is a
nearby dwarf galaxy that experienced a very strong starburst more
than about 4 million years ago and now displays outflows of hot
metal-rich gas (Martin et al.~\cite{martin02}). NGC\,4449 is an
irregular starburst galaxy, which at optical wavelengths has
properties similar to the Lyman-break galaxies (LBG) at high
redshift (Annibali et al.~\cite{annibali08}).

The mean equipartition magnetic field strengths for our sample of LG
dwarfs is $<4.2\pm 1.8\,\mu$G, including the upper limits of the
field strength for the radio-undetected dwarfs at 2.64\,GHz (Table
\ref{t:bfields}). The typical magnetic fields in dwarfs are thus
significantly weaker than in spiral galaxies, for which the mean
field strength is about $10\,\mu$G (Beck~\cite{beck05}). In our LG
sample, the strongest field is observed in the starburst dwarf IC\,10
and its value of $9.7\,\mu$G is close to the that of starburst dIrrs
(NGC\,1569 and NGC\,4449) from outside of the LG. Both Magellanic
Clouds have magnetic fields close to the estimated mean value for
our LG sample.

We also estimated the strength of the {\em ordered} magnetic field
of the LG dwarfs for which polarized emission was detected at
4.85\,GHz ($4.5\pm0.5$, $5.1\pm2.1$, and $0.10\pm0.04$\,mJy for
IC\,10, NGC\,6822, and IC\,1613, respectively). The strengths are in
the range of $0.4-0.9\,\mu$G. The ordered-to-random field ratio of 
those galaxies is as large as about 0.2, which is similar to the averaged
field order in spiral galaxies. For undetected dwarfs, we expect 
the production of an ordered field to be of the same efficiency or
lower.

\section{Discussion}
\label{s:discussion}

As presented in Sect.~\ref{s:magnetic}, the magnetic fields in LG
dwarfs are statistically almost three times weaker than in typical
spiral galaxies of various kinds. To estimate the influence
of the star formation rate and other galactic properties on the
magnetic field production, we compiled various characteristics of
our dwarfs available in the literature. Table \ref{t:bfields}
includes global SFRs, \ion{H}{i} masses ($\rm{M_{HI}}$), infrared
fluxes at $60\,\mu$m, and rotational velocities of all dIrrs in 
our sample. Knowing the global SFR, we calculated the SFR surface density
($\Sigma$SFR) using object sizes from Table \ref{t:objects}. We obtained 
the gas surface density $\Sigma\rho$  from the disk-averaged
\ion{H}{i} atomic gas. Since in dwarf galaxies, the
molecular gas contributes to the total gaseous mass only up to a few
percent for the most massive objects (see e.g. Braine et al.~\cite{braine01}), 
the atomic component closely represents the total gaseous mass of dwarfs.

Below, we analyse various correlations of magnetic field
with other dwarf characteristics. For the sake of comparison we also
analyse four other well-known irregular dwarfs: both of the Magellanic Clouds, NGC\,1569,
and NGC\,4449.

\begin{figure*}[t]
\begin{center}
\includegraphics[width=6.3cm]{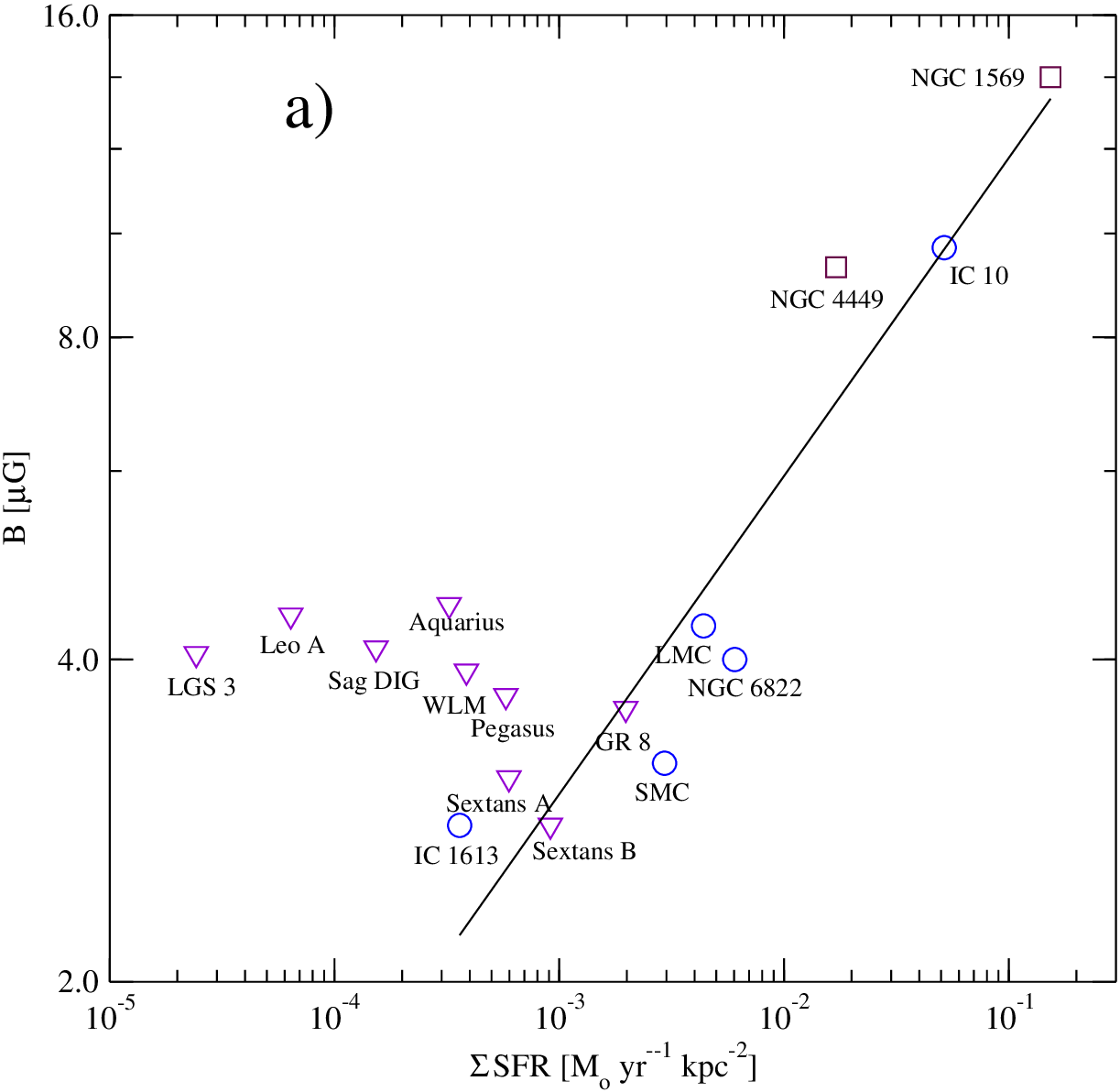}
\hspace{0.9cm}  \includegraphics[width=6.4cm]{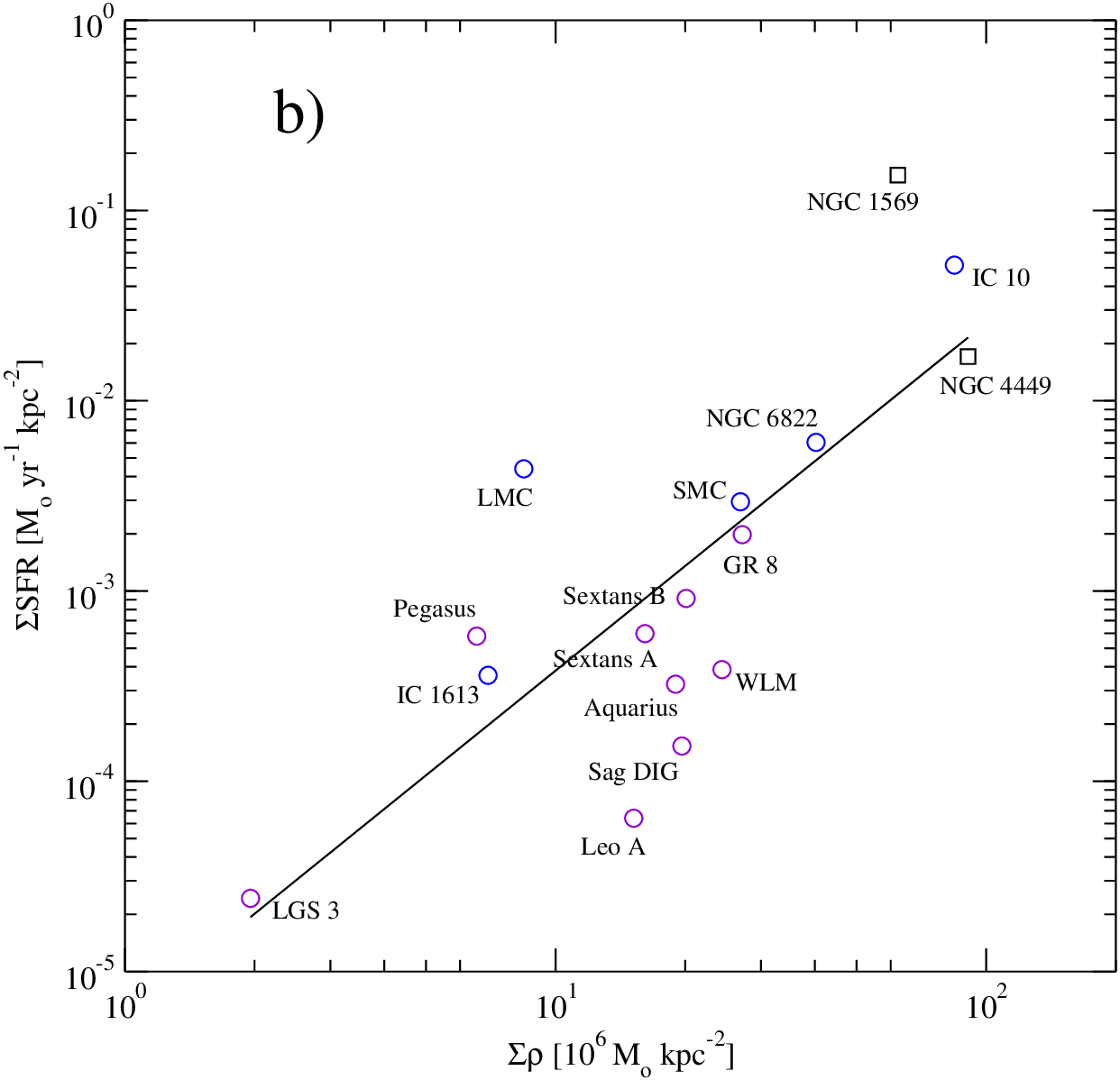}
\includegraphics[width=6.3cm]{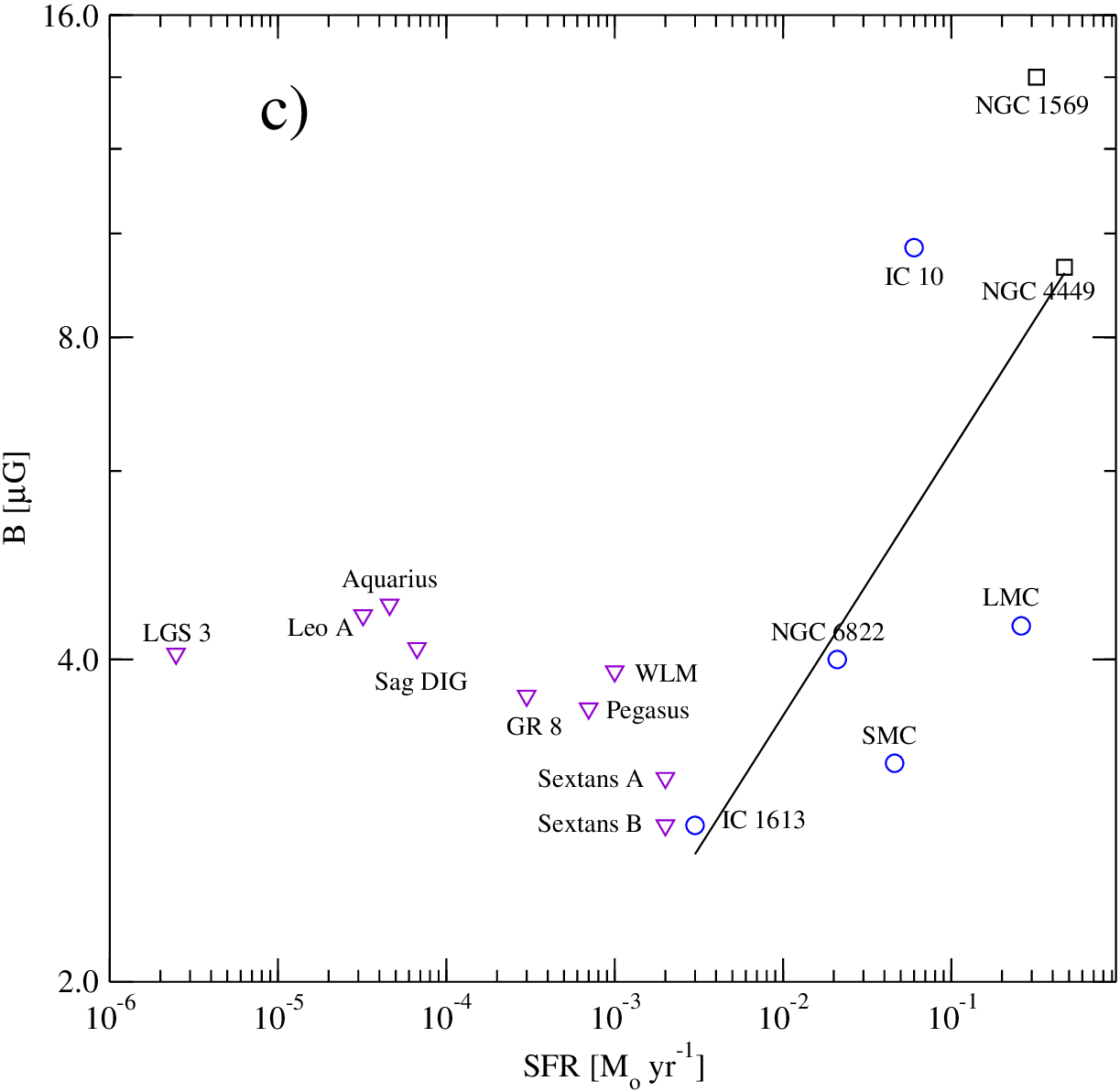}
\hspace{1cm}  \includegraphics[width=6.5cm]{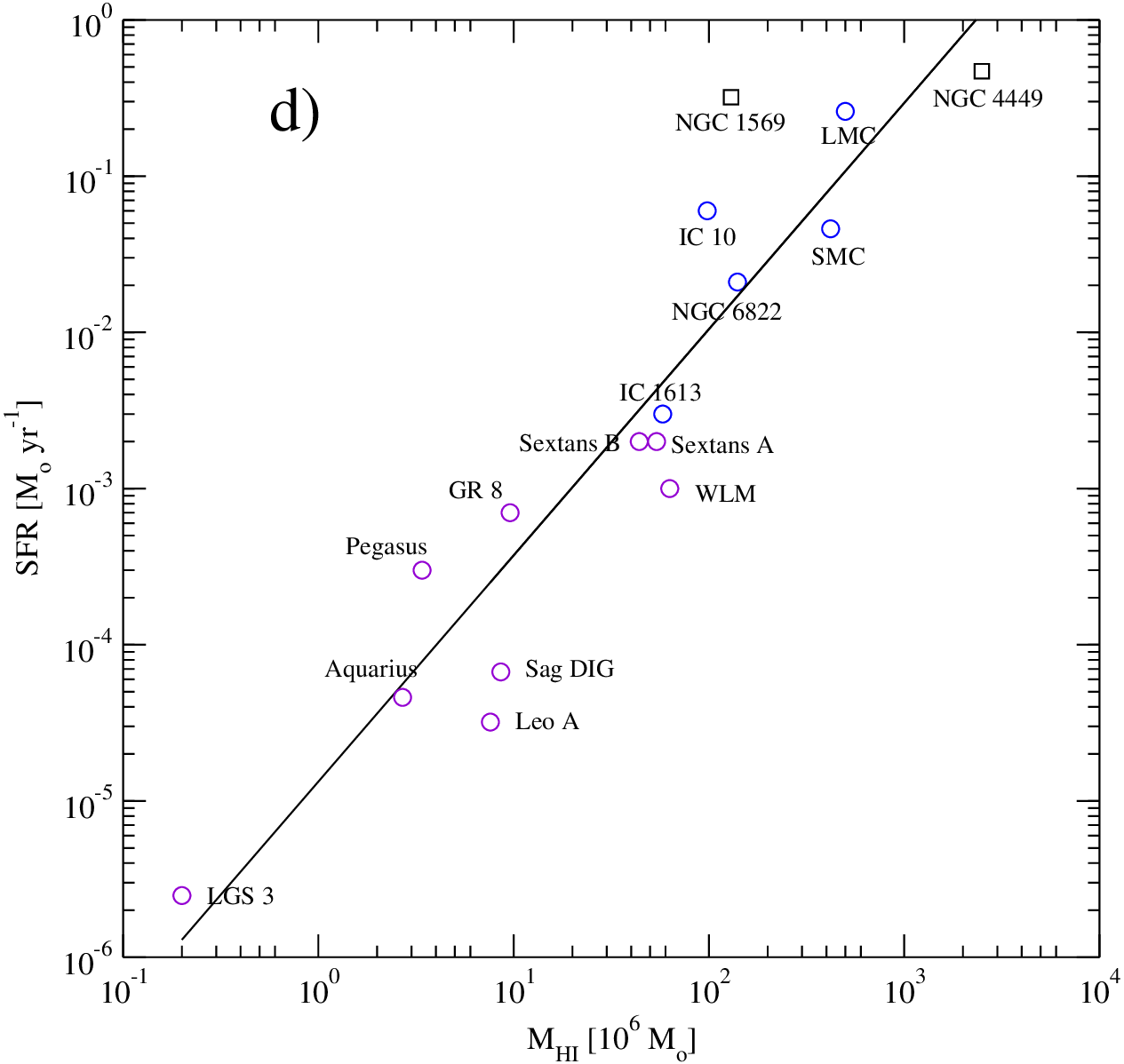}
\end{center}
\caption{
Correlations of the magnetic field strength and SFR with other parameters. LG
dwarfs from our sample, LMC, and SMC are marked by circles or in cases when
the upper limit of the radio emission was applied -- by triangles. Comparison
dwarfs are marked by squares. The solid lines present power-law fits. Fits a and c
are restricted to radio-detected dwarfs only (circles and squares).
}
\label{f:corel}
\end{figure*}

\begin{table*}[t]
\caption{Correlation coefficients, their significance levels in
percentages, and the number of dwarfs included in calculations.}
\begin{center}
\begin{tabular}{ccccccc}
\hline
\hline
            &$\Sigma$SFR & $\rm{M_{HI}}$ & $\Sigma\rho$ & $L$       & $B$    \\
\hline
SFR          & 0.90 (0.001\%;16) & 0.93 (0.001\%;16)  & 0.65 (1\%;16) & 0.82 (0.01\%;16) &0.70 (8\%;7)\\
$\Sigma$SFR &   $-$       & 0.72 (0.2\%;16)  & 0.77 (0.001\%;16) & 0.50 (5\%;16) &0.94 (0.2\%;7)\\
$\rm{M_{HI}}$&   $-$       &    $-$       & 0.66 (1\%,16) & 0.92 (0.001\%:16) &0.16 (73\%;7)\\
$\Sigma\rho$ &   $-$       &    $-$       &   $-$       & 0.33 (21\%:16)&0.78 (4\%;7)\\
$L$          &   $-$       &    $-$       &   $-$       &   $-$       & $-0.36$ (43\%,7)\\
\hline
\end{tabular}
\end{center}
\label{t:corel}
\end{table*}

\subsection{Main factors regulating magnetic field}
\label{s:sfr}

We determined mutual correlations between a range of parameters: magnetic
field strength $\vec{B}$, $SFR$, $\Sigma SFR$, $\rm{M_{HI}}$, $\Sigma\rho$, and the
dwarf's unprojected linear size $L$, taken as the source major axis from
Table~\ref{t:objects}. Where possible, we included in the calculations
all 16 dwarfs, including the comparison dwarfs. When searching for correlations
with the magnetic field strength, we restricted our calculations to 
radio-detected dIrrs (seven objects). Owing to small number statistics, we
chose to evaluate the Pearson correlation coefficient $r$ and in each case 
performed a test of its statistical significance. We determined
the significance level as the probability of rejecting a hypothesis
that $r=0$. Results are presented in Table~\ref{t:corel}. In a
similar way, we also performed analogous calculations for
the Kendal rank correlation coefficients. This approach led to the
same following conclusions.

We found that among the various relations studied the magnetic
field strength depends primarily on the density of the star
formation rate $\Sigma$SFR having the largest correlation
coefficient $r=0.94$ (significant at the $P=0.2\%$ level). This 
strong dependence can be intuitively understood because the local SFR determines
the population of supernova explosions, which constitute the main
source of the turbulent energy, which is in turn vital to the
dynamo process (Arshakian et al.~\cite{arshakian09}, see also
Sect.~\ref{s:dynamo}). This relation shown in Fig.\,\ref{f:corel}a
can be quantified by the power-law fit $\vec{B}\propto
\Sigma$SFR$^{0.30\pm0.04}$.
A similar nonlinear relation between total magnetic field strength
and the global SFR has been observed for nearby spiral
galaxies (Krause~\cite{krause09}) and fits to the equipartition model for the
radio-FIR correlation (Niklas \& Beck~\cite{niklas97}).
A strong influence of the local SFR on the (random) magnetic field has also been 
observed {\em within} the disk of a large spiral galaxy
NGC\,4254 (Chy\.zy~\cite{chyzy08}).

The remaining group of radio-undetected dwarfs occupy a
common region in the $\vec{B}-\Sigma$SFR plane. If detected, they would possibly
move down in this plane as their current positions represent only
upper limits to the magnetic field strength. Therefore, the
population of undetected dwarfs would likely fulfil the power-law
determined for the radio-detected dwarfs alone. We give
another argument for this prediction in Sect.~\ref{s:radiofir}. We
also notice a gap in Fig.\,\ref{f:corel}a between dwarfs of weak and
strong magnetic fields. This may be evidence of a 
threshold in either the dynamo action producing magnetic fields,
or the SFR above which
($\Sigma$SFR$\approx10^{-2}$\,M$_{\sun}$ yr$^{-1}$ kpc$^{-2}$) the
dynamo activity is enhanced.

The physical processes underlying both the star formation and magnetic fields in our
sample of dwarfs are usually modelled by relating the surface
density of galactic SFR ($\Sigma$SFR) to the gas surface
density $\Sigma\rho$ (Kennicutt~\cite{kennicutt98}). The empirical
fit to a power-law relation $\Sigma{\mathrm SFR}\propto\Sigma\rho^N$ with $N\approx
1.4$ was found to provide an appropriate parametrization of these
processes for different types of galaxies (Schmidt~\cite{schmidt59}). 
It follows that the gas density is a
major factor influencing the star formation rate. We find that our
dwarf galaxies also follow a similar power-law fit
(Fig.\,\ref{f:corel}a, Table~\ref{t:corel}) with a slope of
$N=1.83\pm0.30$ and a correlation of $r=0.77$ (which is significant at much
less than the $P=1\%$ level). The slope is somewhat steeper
but agrees within its broad uncertainty limits with the aforementioned
Schmidt law. The slightly steeper relation found for dwarfs,
which are relatively low-mass objects, is also reasonable, as some
star-formation thresholds in low-density (low-mass) galaxies are
expected (Kennicutt~\cite{kennicutt98}). It is also likely that the
Schmidt law is a strong oversimplification as \ion{H}{i} gas is only
a weak tracer of star formation, and even CO gas has a too low
excitation temperature to serve as a good star formation tracer.
Hence, a significant fraction of the gas is unrelated to star
formation, explaining the loose correlation shown in
Fig.\,\ref{f:corel}b.

\begin{figure}
    \begin{center}
        \includegraphics[width=7.3cm]{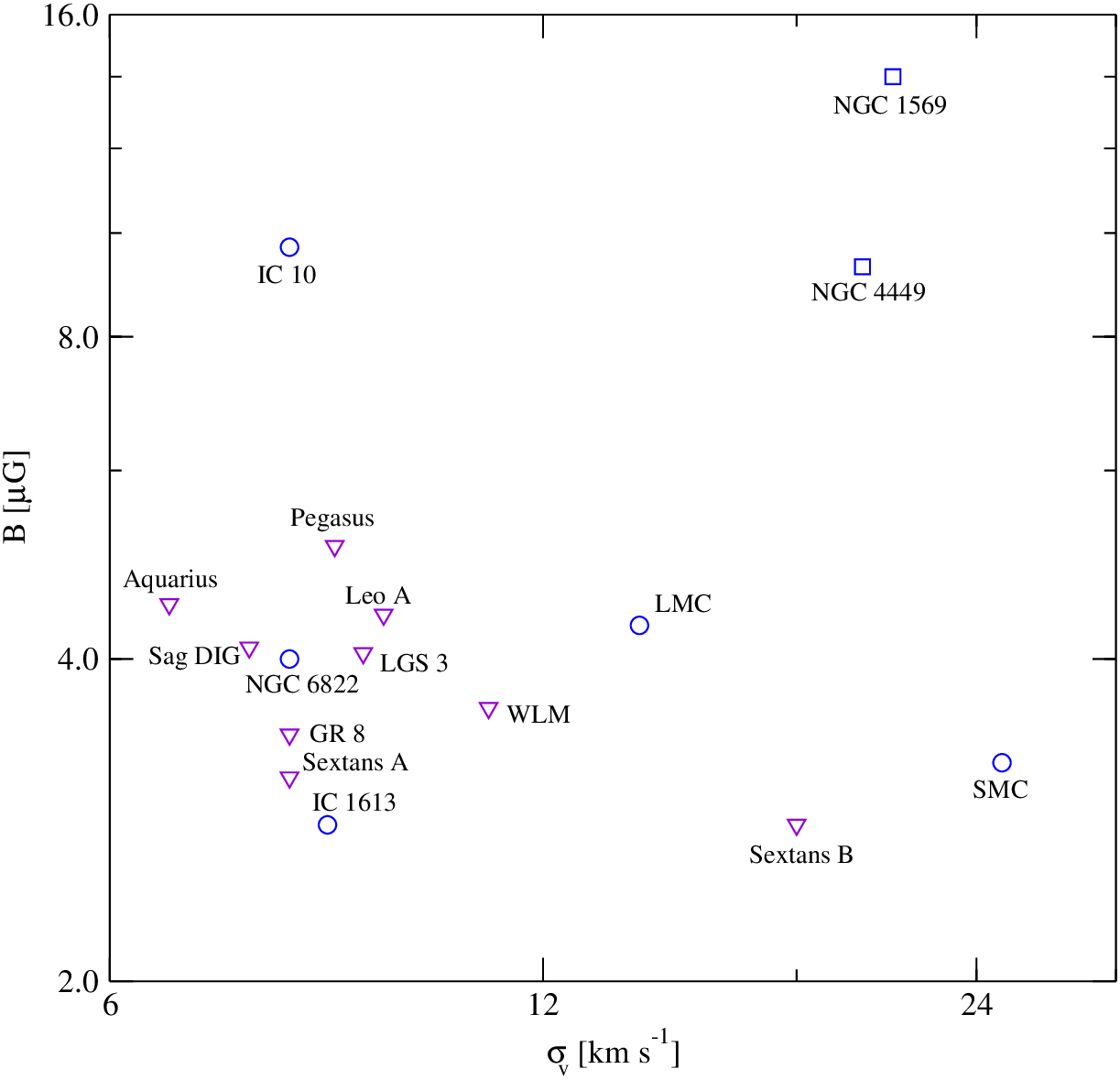}
    \end{center}
    \caption{
Magnetic field strength versus velocity dispersion for dwarf galaxies.
See Fig. \ref{f:corel} for symbol coding.
}
\label{f:sigma}
\end{figure}

The $\Sigma$SFR -- $\Sigma\rho$ relation observed for our dwarfs
causes in turn a significant correlation for the gas density and the
magnetic field strength (Table~\ref{t:corel}). Hence, in contrast 
to the $\vec{B}-\Sigma$SFR relation, we can also describe the
magnetic field in dwarfs as being controlled by the local gas
density. For the radio-detected dwarfs, we obtained the relationship
of $\vec{B}\propto \Sigma\rho^{0.47\pm0.09}$ with a correlation of
$r=+0.78$ ($P=4\%$). This relation is very close to that determined
for spiral and irregular galaxies by Niklas \& Beck~(\cite{niklas97}), 
who found an exponent of $0.48\pm0.05$.

It was shown that for a wide range of galaxies the far-infrared
luminosity is a linear indicator of the SFR (Kennicutt~\cite{kennicutt98}). 
Thus, to test this relation for dwarfs we can
estimate $\Sigma$SFR for all investigated dwarfs independently of the 
H$\alpha$ emission using the available infrared $60\,\mu$m
fluxes (Table \ref{t:bfields}). After determining the dependence of
gas density on star formation (exponent 1.83) and the relation
of magnetic field to gas density (exponent 0.47), we can model the
relation between magnetic field and the surface-normalized infrared
luminosity of $\vec{B}\propto \Sigma L_{IR}^{0.26}$. The {\em
observed} relation for radio-detected dwarfs gives a power-law fit
with the exponent of $0.25\pm 0.04$ (correlation $r=+0.90$,
$P=0.5\%$), which is in a good agreement with the predicted one. We
also found that this relation is compatible with the one determined
for the $\Sigma$SFR estimated in a similar way from H$\alpha$
emission (with the slope $0.30\pm0.04$). In summary, we can state
that in dwarf galaxies the gas density regulates both the
star-formation rate and the magnetic field production in a way
similar to the spiral galaxies (Krause~\cite{krause09}) and the 
late-type galaxies (Chy\.zy et al.~\cite{chyzy07}) studied so far.

We also found that the magnetic field is stronger in dIrrs of
higher {\em global} SFR (Fig.\,\ref{f:corel}c). This trend for
the seven radio-detected dwarfs can be quantified by the relation
$\vec{B}\propto$SFR$^{0.25\pm0.06}$ with the correlation coefficient
of $r=+0.70$ ($P=8\%$). A similar relation with an exponent of
$0.34\pm0.08$ was fitted for the spiral galaxies by Niklas \& 
Beck~(\cite{niklas97}), while Vall\'ee~(\cite{vallee04}) obtained an
exponent of $0.13\pm0.04$ for a different sample of nearby
spirals. In our estimate, the exponent could be just a lower limit,
as we took into consideration only the radio-detected (hence
possibly relatively brighter) dwarfs (see Fig.\,\ref{f:corel}c). The
correlation found between the {\em global} SFR and $\vec{B}$ must
involve some additional factors, not accounted for by the local
$\vec{B}-\Sigma$SFR dependency discussed above
(Fig.\,\ref{f:corel}b).

One possible explanation of the $\vec{B}-$SFR relation could be the
galactic mass. According to the principal component analysis of various global
characteristics of galaxies performed by Disney et al.~(\cite{disney08}),
the galactic mass is the principal component strongly affecting 
a wide range of other global galactic properties. Plotted together, the
\ion{H}{i} mass and SFR of dwarfs are nonlineary related by a power-law slope
of $1.45\pm0.11$ ($r=+93$, $P\ll1\%$, Fig.\,\ref{f:corel}d).
However, we do not observe a statistically significant influence of the
mass on magnetic field strengths in dwarfs ($r=+0.16$, $P=73\%$). A larger sample
of radio-detected dwarfs is certainly needed to confirm this finding
at higher statistical confidence.

If star formation in LG dwarfs were also a driver of effective 
galactic winds (see Sect. \ref{s:model}), then one might expect 
weaker magnetic fields in actively star-forming dwarfs because 
the fields and CRs could just escape in galactic outflows. 
To check this idea, we compare the strength of 
the magnetic field $\vec{B}$ with the velocity dispersion $\sigma_v$ of ISM, 
which should scale with the star formation activity. In the case of effective winds, 
larger velocity dispersion should correspond to weaker fields.
We present the values of $\sigma_v$ estimated from 
\ion{H}{i} observations in Table \ref{t:bfields}, and the derived 
$\vec{B}-\sigma_v$ relation in Figure \ref{f:sigma}. 
The dIrrs with starbursts (NGC\,1569, NGC\,4449) have large velocity 
dispersions ($\sigma_v>20$\,km\,s$^{-1}$) but also have strong magnetic 
fields ($\vec{B}>10\,\mu$G). The larger $\sigma_v$ of both these galaxies 
and the Magellanic Clouds could be caused by galactic winds and/or by specific 
gas motions due to their tidal interactions. 
For dIrrs with weak magnetic 
fields ($\vec{B}<5\,\mu$G), the velocity dispersion is around 
$10\pm2$\,km$^{-1}$, which is a value typically found in quiescent galaxies.
Therefore, to explain the $\vec{B}-\sigma_v$ relation of dIrrs, galactic 
outflows are not needed, hence we ascribe the presence of weak 
magnetic fields to low star formation and not to galactic winds.

A possible process that could also influence both the SFR and magnetic fields is
the gravitational interaction of galaxies. Young et al.~(\cite{young96}) found
that the efficiency of star formation in interacting galaxies is significantly
higher than in isolated objects. For the analysed dIrrs, clear signs
of gravitational perturbations or gas infall were clearly detected in
all most-intensively star-forming objects: NGC\,4449, NGC\,1569, IC\,10,
LMC, and SMC (see Table \ref{t:bfields}). Therefore, in dIrrs gravitational
interactions can indeed stimulate massive star formation and promote stronger
magnetic fields in a way similar to that observed in the Antennae galaxies
(Chy\.zy \& Beck~\cite{chyzy04}). A larger sample of radio detected dIrrs is again
necessary to confirm this possibility.

\subsection{SFR evolution}
\label{s:evolution}

\begin{figure}
    \begin{center}
        \includegraphics[width=7cm]{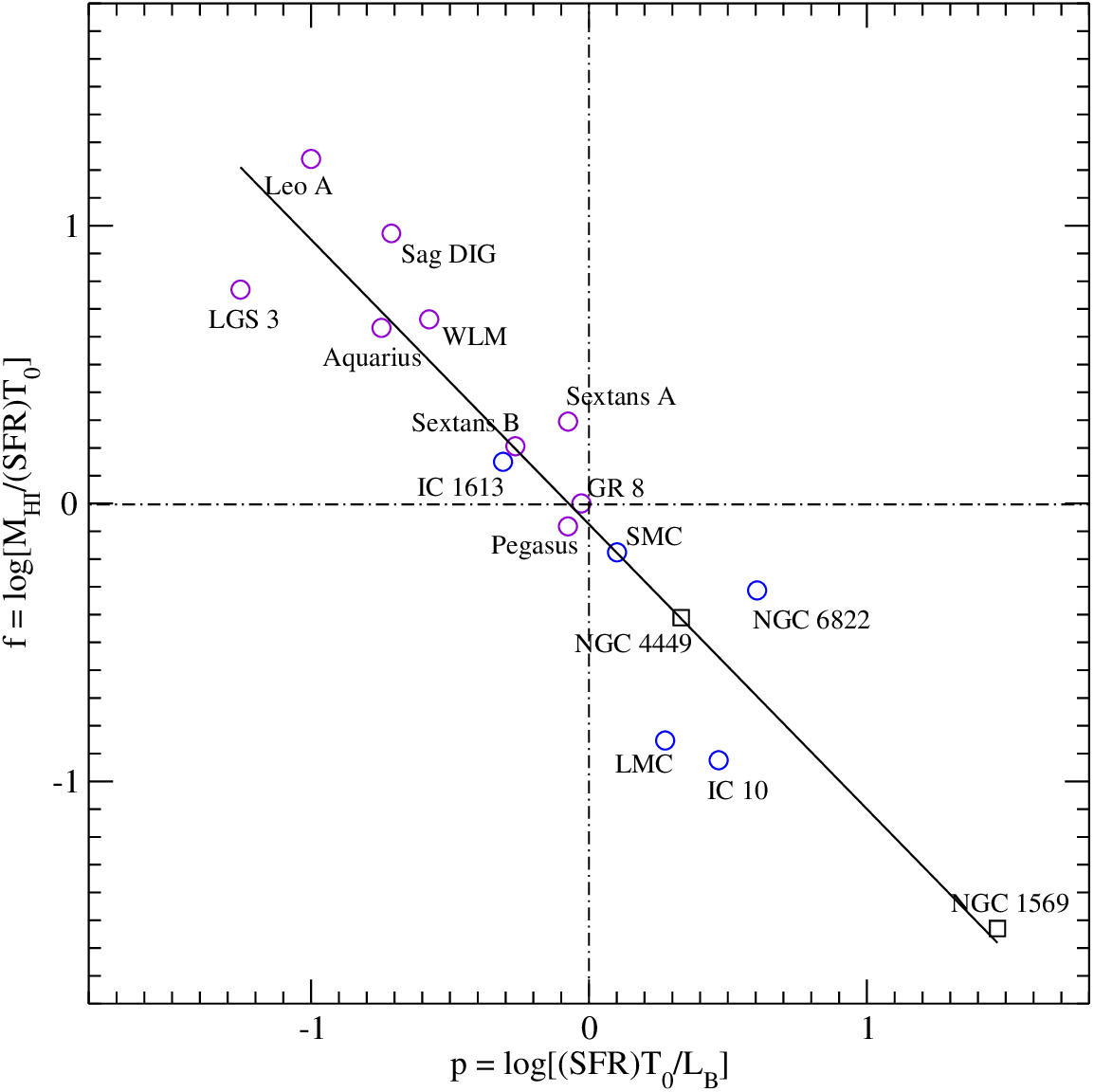}
    \end{center}
    \caption{
Dwarf galaxies on the past-future evolution plane: $p={\rm log}({\rm SFR}\,\,T_0/L_B)$ and
$f={\rm log}[M_{HI}/({\rm SFR}\,\,T_0)]$.
The comparison galaxies from outside of the Local Group are shown by squares.
}
\label{f:p-f}
\end{figure}

The magnetic field in dIrrs, as in normal spiral galaxies, could
be related not only to the current state of star-forming activity,
as discussed above, but also to their recent star formation history.
The production of magnetic fields might then be connected
to the galaxy past. The evolution of global star formation can be
qualitatively studied in a plane of two dimensionless parameters,
$p={\rm log}({\rm SFR}\,\,T_0/L_B)$ and $f={\rm log}[M_{HI}/({\rm
SFR}\,\,T_0)]$, where $L_B$ denotes the total blue luminosity of a
galaxy and $T_0=13.7$\,Gyr the age of the Universe. The former
parameter ($p$) characterizes the past, and the latter ($f$) the
future of the galactic star formation (see Karachentsev \& 
Kaisin~\cite{karachentsev07} for details). We calculated the values of 
$L_B$ for dwarfs from the absolute B-magnitude given in 
Mateo~(\cite{mateo98}) or, if unavailable, from the LEDA database. The
other parameters needed are given in Table~\ref{t:bfields}. We found
that the investigated LG dIrrs show a similar distribution across
the $p-f$ plane (Fig.\,\ref{f:p-f}) to the group of dwarf galaxies
around the giant spiral M\,81 (Fig. 4 in Karachentsev \& 
Kaisin~\cite{karachentsev07}), thus seem to be representative of our
nearby region of the Universe.

The majority of all dwarfs with detectable radio emission and
magnetic fields are located in the bottom-right quarter of this
plane (Fig.\,\ref{f:p-f}). Despite being more massive objects, they
produce stars so efficiently, that their \ion{H}{i} gas would be
sufficient for just a short period of time, e.g. for 10\% of the
Hubble time ($f\approx-1.0$) for LMC and IC\,10. The most extreme
case among all dwarfs is the starburst NGC\,1569 with $f=-1.5$. Its
star formation, and accordingly also the production of a magnetic
field, should cease soon (the current level of SFR may continue 
only for the next 400\,Myr). Galaxies such as NGC\,1569 are presumably
at their major star-forming phase (large $p$ value of 1.5), thus
might be analogs of starbursting (`young') objects of the early
Universe. This conclusion is supported by an analysis of the star
formation history of NGC\,1569 derived from a synthetic
color-magnitude diagram method incorporating strong episodes of star
formation over the past Gyr (Angeretti et al.~\cite{angeretti05}).
Within our unbiased LG sample the only similar dwarf is IC\,10
($p\approx +0.5$) with strong magnetic fields of $9.7\,\mu$\,G. The 
LMC is located at a similar location in the $p-f$ diagram. Its
star-formation episodes are believed to be triggered by close
encounters with the Milky Way over the past 4\,Gyr (Bekki \& Chiba
~\cite{bekki05}).

In contrast to the aforementioned bottom-right quarter, the top-left
quarter of the $p-f$ plane is occupied by less evolved objects with
a relatively weak star-formation activity (negative $p$) that 
evolve slowly, despite having enough gas to produce stars. Galaxies
such as WLM, Sag DIG, Aquarius, LGS\,3, and Leo\,A ($f\ge 0.5$) could even
make stars for at least three times the Hubble time with the current
SFR. These galaxies are most common among the LG dwarfs and all
lack any detectable magnetic field. Thus, the existence of
detectable magnetic fields in dwarfs is not coincidental but 
connected to the whole galaxy evolution.

To check whether the magnetic properties of dwarfs are indeed
related to their overall evolution we investigated the relation of
magnetic field strength to the mean metallicity (12+O/H) of the
dwarfs, which we obtained for all the objects from the compilation
of Mateo~(\cite{mateo98}) and Braine et al.~(\cite{braine01}). We
found that strong magnetic fields are observed exclusively in the
evolved systems of relatively high metallicity. No object of
metallicity 12+O/H$<$8.1 has a magnetic field stronger than
$5\,\mu$G, which indicates that the magnetic field production in
dwarfs is tightly connected to the galactic gas consumption during
their lives. Since the metallicity of dwarfs strongly correlates
with their masses (Braine et al.~\cite{braine01}), this can explain
the complex and mutual relation of magnetic field strength, global
SFR, and mass (Sect.\,\ref{s:sfr}).

A good example to illustrate these complex issues is the
well-studied NGC\,4449. It shows a global starburst with a
widespread star-formation activity. It forms stars twice as rapidly
as the LMC, even though sizes of both galaxies are similar (Table
\ref{t:bfields}). NGC\,4449 is also unusual in having neutral
hydrogen gas extending to six times its Holmberg radius. The large
amount of \ion{H}{i} gas places this galaxy closer to the center of
the $p-f$ plane than other starburst dwarfs. In the past, the galaxy 
most likely acquired some gas from an interaction or possibly a
merger, as its inner part is counter-rotating with respect to the
outer envelope (Hunter et al.~\cite{hunter98}). The strong
and regular magnetic field of NGC\,4449 (almost $10\,\mu$G) is therefore
likely connected to the star formation history, star-formation
triggering mechanisms, the acquired mass, and the entire complicated
history of the galaxy. The omission of these processes could have prevented a
fully successful numerical MHD modelling of this object
(Otmianowska-Mazur et al.~\cite{otmianowska00}). This example also
shows that dIrrs do not seem to be simple galaxy systems,
nor easier to model than typical spiral galaxies.

\subsection{Magnetization of the IGM}
\label{s:model}

The magnetic field strength defines statistical scaling relations
with other properties of LG dwarfs, such as $\Sigma$SFR, $\Sigma \rho$,
and mass (Sect.~\ref{s:sfr}). Similar trends can also be observed
for local regions within large spiral galaxies (Chy\.zy~\cite{chyzy08}), 
indicating in both cases some universal connections
between various physical mechanisms. Close links between the global
stellar and dynamical properties of dwarfs were also observed by Woo
et al.~(\cite{woo08}) and modelled for higher-redshift dwarfs by
Tassis et al.~(\cite{tassis08}). We therefore infer that the
magnetic properties of LG dwarfs can tell us something about
magnetic fields in dwarf galaxies in past cosmological epochs,
in particular about the role they might have played in magnetizing
the intergalactic medium in the early Universe. Low-mass
galaxies with a weak gravitational potential seem to be good
candidates to support this process. We perform below a
straightforward modelling of whether dwarf galaxies could actually be
an efficient supplier of magnetic fields to the IGM. We use the
available data concerning the magnetic field characteristics of the
low-mass galactic systems that we observed within the Local Group
and our current knowledge of protogalaxies and their environments. 

According to theoretical considerations, very weak seed magnetic
fields of the order of $10^{-20}-10^{-18}$\,G could have been
generated in early cosmological epochs by a number of mechanisms,
including phase transitions, the Weibel instability (Medvedev \&
Loeb~\cite{medvedev99}), and the Biermann battery (Widrow~\cite{widrow02}, 
Zweibel~\cite{zweibel06}). Alternatively, seed
fields could have been produced by the first stars and amplified by
dynamo processes in protogalaxies. The small-scale galactic dynamo
could produce strong magnetic fields on timescales of even
$10^7$\,yr (Brandenburg \& Subramanian~\cite{brandenburg05}, Arshakian et 
al.~\cite{arshakian09}). Outflows from protogalaxies could then seed the
IGM and massive galaxies that assembled at lower redshifts with
magnetic fields (Kronberg et al.~\cite{kronberg01}, Bertone et 
al.~\cite{bertone06}, Donnert et al.~\cite{donnert09}, Kronberg et 
al.~\cite{kronberg99}, Samui et al.~\cite{samui10}, Dubois \& 
Teyssier \cite{dubois09}). Measurements
of integrated Faraday rotation of distant quasars as well as the
modelling of the cosmological evolution of the IGM provide hints of
a possible significant magnetization of the IGM at $z\approx 4-7$
(Kronberg~\cite{kronberg06}). 

\begin{table}[t]
\caption{Modelling of the magnetization of the IGM. From the
assumed values of redshift $z$, $T_{IGM}$, $\epsilon$, $R_0$, $B_0$,
SF mass or SFR, we first model $E_w$ by the Starburst99 code and
next obtain $R_s$ and $B_s$ (see the text for details). }
\begin{center}
    \begin{tabular}{lcccccc}
\hline
\hline
Type              & Pri dSph & Pri dIrr & LBG & LBG \\
     & \multicolumn{4}{c}{instantaneous star formation} \\
\hline
SF Mass                 & 2.0e5 & 1.0e7 & 4.0e8 & 4.0e9 \\
Redshift $z$            & 8 & 7 & 5 & 3 \\
Wind energy $E_w$ [erg] & 4.0e54 & 2.0e56 & 8.0e57 & 8.0e58\\
SF size  $R_0$ [kpc]    & 0.2 & 0.7 & 1.0 & 2.0\\
Stall radius $R_s$ [kpc]& 9 & 36 & 163 & 528 \\
$B_0$ [G]               & 1.0e-6 & 1.0e-5 & 2.3e-5 & 4.0e-5\\
$B_s$ [G]               & 5.3e-10 & 1.9e-9 & 8.6e-10 & 5.7e-10 \\
\hline
\\
Type & \multicolumn{4}{c}{Local Group dIrrs} \\
     & \multicolumn{4}{c}{continuous SF} \\
\hline
SFR                     & 0.00001& 0.0003 & 0.01  & 0.1 \\
Redshift $z$            & 0  & 0 & 0 & 0 \\
Wind energy $E_w$ [erg] & 3.0e50 & 1.5e52 & 3.0e53 & 3.0e54 \\
SF size  $R_0$ [kpc]    & 0.05 & 0.2 & 0.4 & 0.7\\
Stall radius $R_s$ [kpc]& 0.2 & 0.9 & 2.3 & 5.0 \\
$B_0$ [G]               & 5.0e-7 & 1.0e-6 & 3.0e-6 & 8.0e-6 \\
$B_s$ [G]               & 2.3e-8 & 5.5e-8 & 8.8e-8 & 1.5e-7 \\
\hline
\end{tabular}
\label{t:model}
\end{center}
\end{table}

On the other hand, the observations of Ly-$\alpha$ quasar absorption
lines indicate that metals are present in the IGM up to $z\approx5$
(Cowie et al.~\cite{cowie95}; Songaila \& Cowie~\cite{songaila96}; 
Ellison et al.~\cite{ellison00}, Bouch{\'e} at al. \cite{bouche07}). 
Various estimates suggest that clouds obscuring quasars are 
located 200-500\,kpc from protogalaxies (Meiksin~\cite{meiksin09} 
and references therein). To transport metals into these regions, 
some mechanisms, such as stellar winds, supernova-driven galactic
outflows, or AGN induced outflows, must be at work there. The same
galactic outflows that pollute the space with metals could also
supply the IGM with magnetic fields produced in protogalaxies.

In our modelling, we study a galactic-wind-blown bubble which energy
$E_b$ drives the bubble expansion (Veilleux et al.~\cite{veilleux05}, 
Meiksin~\cite{meiksin09}). Only some net fraction $\epsilon$ of the injected
energy $E_w$ from supernovae and stellar winds is available for this
process ($E_b=\epsilon E_w$), which can only be roughly estimated to
be within $0.01-0.1$, as it depends on several unknown factors such as:
expansion losses, wind mass loading, radiative cooling, uniformity
of the ambient medium, and energy losses intrinsic to the
interstellar medium (e.g. Bertone et al.~\cite{bertone06}, Cho \&
Kang~\cite{cho08}, Meiksin~\cite{meiksin09}). The bubble of radius
$R_b$ and thermal pressure $P_b=E_b/2\pi R^3_b$ finally reaches an
equilibrium with the IGM pressure $P_{IGM}\propto T_{IGM} (1+z)^3$
and stops expanding at the stall radius $R_s=(E_b /
T_{IGM})^{1/3} (1+z)^{-1}$.

Next, we assume that magnetic fields are also blown out along with
the plasma that has escaped from star forming regions and reached the strength
$B_s$ at radius $R_s$. For spherical expansion of the bubble, the
strength of the magnetic fields dominated by the random component
scales with the bubble's density as $B_b\propto\rho^{2/3}$.
Following the results of Sect.~\ref{s:sfr}, we relate the initial
strength of the magnetic fields $B_0$ within a star forming region
of extent $R_0$ to the total galactic mass and the global SFR, using
masses and sizes of the protogalaxies estimated from observations
(see below) (Table \ref{t:model}).

We calculated the amount of mechanical energy $E_w$ injected by active
star-forming regions by modelling an evolutionary stellar population 
synthesis using the Starburst99 
code\footnote{The code used is the latest version 5.1 (April 2006)}
(Leitherer et al.~\cite{leitherer99}, Vazquez \& 
Leitherer~\cite{vazquez05}). The main input
parameter of this code is the star forming (SF) mass in the case of
instantaneous star formation, or the SFR in the case of continuous
star formation. For high-z objects, we used the former option to
simulate the burst of star formation during the formation process of
primordial galaxies, as well as the enhanced Geneva stellar tracks
of low metallicity $Z=0.004$, appropriate for young galaxies. For
the stellar initial mass function, the classical Salpeter law was used
in-between 1 and 100\,M$_{\sun}$.

According to the CDM scenario, the first dark matter halos appeared in the
Universe at $z\approx 20-8$. The observational characteristics of
dSphs in the LG (see Sect.\,~\ref{s:sample}), which have the common
total mass of about $10^7$\,M$_{\sun}$ (Strigari et al.~\cite{strigari08}),
suggest that they may constitute LG fossils of primordial dSph (Pri
dSph) galaxies (Ricotti~\cite{ricotti10}). Almost all LG dSphs
exhibit a prominent stellar population of about 10 Gyr in 
age, resulting from a SFR of about several hundred M$_{\sun}$ per Myr
(Dolphin et al.~\cite{dolphin05}) followed by a subsequent
decline in SF activity. The lower-mass galaxies, which fossils
can now be observed as UF dSphs, were less efficient in forming stars 
and transformed a smaller fraction of their baryonic mass into stars
(Salvadori \& Ferrara~\cite{salvadori09}). In our
modelling, we therefore restricted the low-end of the instantaneous SF mass to
$2\times10^5$\,M$_{\sun}$ (Table \ref{t:model}). For a total
galactic mass of dSphs of e.g., several  $10^7$\,M$_{\sun}$, this
corresponds to 14\% of the baryonic content, converting 3.5\% of the
gas into stars in a single outburst. The applied values of SF mass
correspond to a mean SFR of $5\times10^{-3}$\,M$_{\sun}$yr$^{-1}$
over the time of the starburst ($\approx 4\times10^7$\,yr), which is
close to the SFR estimated by Dolphin et al.~(\cite{dolphin05}) and several
orders of magnitude higher than the SFR observed today among LG
dSphs. They are also similar to those used in simulations of dwarf
galaxy evolution by Ricotti et al.~(\cite{ricotti08},~\cite{ricotti10}).

Low-mass systems might have merged in the past to form
larger galaxies, as predicted by the model of hierarchical
cosmology. The primordial dwarf irregular galaxies (Pri dIrr) 
probably formed later than Pri dSphs or in more massive halos (Ricotti \&
Gnedin~\cite{ricotti05}). In our modelling, we assigned them a SF mass
of $10^7\,M_{\sun}$ (Table \ref{t:model}). This value corresponds to
the case of a strong burst of star formation of about
0.3\,M$_{\sun}$yr$^{-1}$ that can be observed currently in the
starbursting dwarf NGC\,1569 (Table \ref{t:bfields}). This galaxy
was much quieter in the past, including its youth (Angeretti et 
al.~\cite{angeretti05}). Most LG dIrrs probably experienced less
massive bursts of star formation during their evolution. For
example, the SFR mentioned above is two orders of magnitude higher
than the current SFR of IC\,1613 (Table \ref{t:bfields}) and about
one order of magnitude higher than during the most active stage in
the whole evolution of this object (Skillman et al.~\cite{skillman03}).

More massive and starbursting galaxies have been detected as
Lyman-break galaxies (LBGs) (Verma et al.~\cite{verma07}). These
objects are probably progenitors of the present-day early
Hubble-type galaxies and bulges of massive galaxies. LBGs evolved with time in
mass: those assembled at $z\approx5$ are 10 times less massive and
luminous than $z\approx 3$ LBGs (Verma et al.~\cite{verma07}). Our
modelling includes both types of LBGs (at z=5 and z=3) as the
high-end mass systems (Table~\ref{t:model}). The applied SF masses
($4\times 10^8$ and $4\times 10^8$\,M$_{\sun}$) correspond to an equivalent SFR
during the starburst of 10 and 100\,M$_{\sun}$yr$^{-1}$,
respectively. Thus, the second class of LBGs corresponds in terms of
their properties to the M\,82-like starburst galaxies. In galaxies
with masses significantly higher than $10^9$\,M$_{\sun}$, the escape of 
gas from the galactic disks becomes problematic because of the deep
gravitational potential wells (Ferrara \& Tolstoy~\cite{ferrara00}).
Therefore, the modelled $R_s$ for LBGs at z=3 should be regarded as
upper limits. We also note that sometimes even
relatively massive galaxies with a total mass of up to
$\approx10^{11}$\,M$_{\sun}$ are called dwarf galaxies (e.g.
Crain et al.~\cite{crain09}) in the sense that they have a
sub-galactic mass value compared to the typical galactic
systems in the present Universe, such as the Milky Way. However,
primordial systems of this mass, as well as LBGs, would not have
evolved to become the typical LG dIrrs that we investigate in this paper.

For all kind of objects, we adopt in our modelling $T_{IGM}=10^4$\,K
(Meiksin~\cite{meiksin09}) and $\epsilon=0.01$, for which we get for
LBGs at z=3 the bubble's stall radius of $R_s=530$\,kpc
(Table~\ref{t:model}), in agreement with the modelling of metal
enrichment by Madau et al.~(\cite{madau01}), Calura \& 
Matteucci~(\cite{calura06}), and Samui et al.~(\cite{samui08}). Our modelling
predicts a small stall radius of 9\,kpc for primordial dSphs and
36\,kpc for primordial dIrrs (Table~\ref{t:model}). Thus, it is quite
unlikely that the metal absorption systems seen in the Ly-$\alpha$
forest out to at least 300\,kpc from parent objects were produced by
low-mass dIrrs.

If these attempts to explain the IGM's metal enrichment are valid,
our modelling indicates that the primordial dIrrs could have
magnetized the IGM only locally, out to about a 40\,kpc distance, with
the strength of magnetic fields of about a few nanogauss (Table
\ref{t:model}). Accordingly, the contribution from these galaxies is
simply insufficient to have any significant impact on
magnetization of the IGM. Our modelling suggests that more massive
galaxies, such as LBGs, are more effective in magnetizing the IGM, providing a
larger spread (160-530\,kpc) and magnetic field strengths of almost
one nanogauss. Kinematic signatures of vigorous large-scale winds
have been detected among LBGs at $z\approx3-4$ (e.g. Veilleux et
al.~\cite{veilleux05}).

We now examine the effect of varying the model 
parameters. Raising the energy conversion fraction $\epsilon$ from
0.01 (applied in Table~\ref{t:model}) to 0.1 doubles the bubble's stall 
radius, while decreasing the magnetic field strength by factor of five. Thus, the applied
$\epsilon=0.01$ results in upper limits of $B_s$ presented in
Table~\ref{t:model}. Extending the lower end
of the Salpeter IMF to 0.1\,M$_{\sun}$ or raising its upper level to
120\,M$_{\sun}$ results in an input wind energy that is lower by a factor of 2.5
and in turn, a 35\% smaller $R_s$ and 1.8 times stronger magnetic
fields. If we decrease $R_0$ by a factor of two, we obtain a four times weaker
magnetic field. Altering the other input parameters has a less significant effect; 
for example, by changing the IMF to the standard
Kroupa one decreases the input energy by only 68\%. Changing the 
metallicity from 0.004 to 0.02 gives a wind energy higher by 20\%, a
stall radius larger by 7\%, and a magnetic field strength lower by
15\%. A higher star-forming mass or lower IGM temperature by a factor of two 
yield a stall radius larger by about 25\% and a magnetic
field strength weaker by about 60\%. This analysis demonstrates that
the obtained results are most sensitive to the uncertain value of
$\epsilon$. However, even assuming that its uncertainty is as large as
about one order of magnitude does not affect our conclusion
concerning the weak effectiveness of dwarf galaxies in polluting the
IGM with metals and magnetic fields.

We also perform similar modelling of spreading the magnetic fields
of nearby ($z=0$) dwarfs. As LG dIrrs appear to experience continuous star formation
with amplitude variations of factors $2-3$ during their lifetime
(e.g. Grebel~\cite{grebel04},~\cite{grebel05}), we applied this option to
our modelling of stellar population synthesis over a characteristic
timescales of $10^7$\,yrs, and applied a metallicity of $Z=0.02$
(Table \ref{t:model}). Continuous star formation agrees
with the detection of significant amounts of gas in these systems
(Mateo~\cite{mateo98}). We adopted a higher ambient gas
temperature ($T_{IGM}=10^5$\,K), as the IGM in the LG is not pristine
and likely to have a hot component due to galactic feedback, structure
formation heating, and the large potential wells of massive spirals
(Dav\'e \& Oppenheimer~\cite{dave07}, Crain et al.~\cite{crain09}). There
are also some predictions that the density of the local IGM is 
about $10^{-4}$\,cm$^{-3}$ (Pildis \& McGaugh~\cite{pildis96},
Rasmussen et al.~\cite{rasmussen03}, Sembach~\cite{sembach06}), which
we use to estimate the ambient pressure $P_{IGM}$.

Our modelling indicates that typical LG dwarfs do not appear 
capable of providing an efficient supply of magnetic fields to
their environments. Typical wind-blown expanding bubbles may reach
relatively short distances of about 1\,kpc (Table~\ref{t:model}).
Only around the most starbursting dwarfs (as e.g. IC\,10, NGC\,1569,
NGC\,4449) could the IGM be magnetized in this way up to about
0.1\,$\mu$G and within a distance of about 2-5\,kpc.

In the above modelling, the galactic outflows were driven exclusively 
by thermal pressure. However, the energy of CRs may also play an important role in triggering 
galactic winds e.g. Breitschwerdt et al. (\cite{breitschwerdt91}, 
\cite{breitschwerdt93}), Breitschwerdt (\cite{breitschwerdt08}), 
Everett et al. (\cite{everett08}, \cite{everett10}).
This source of energy could be highly important in quiescent 
galaxies, such as our Milky Way (Everett et al. \cite{everett10}). 
Nevertheless, we attempt to estimate the possible influence
of additional pressure from CRs to blow-out bubbles in our low-mass objects.
Treating CRs hydrodynamically and assuming that their pressure could 
reach approximate equipartition with the thermal pressure $P_b$, we repeated the modelling of 
the expansion of the bubble for primordial dSphs and dIrrs. We obtain stall radius of 
11 and 45\,kpc, respectively. These values are only 26\% larger than
in the case of purely thermally driven winds. These estimates seems to 
be upper limits as we do not include any quenching effect of CRs on 
star formation (Socrates et al. \cite{socrates08}, Samui et al. \cite{samui10}) 
and the interaction of CRs with mass-loaded gaseous outflows.

Thus, we conclude that it is highly unlikely that typical dwarf
galaxies (e.g. IC\,1613) could have had any major role in
magnetizing the Universe at any cosmological epoch. Our
predictions of the magnetized surroundings of {\em nearby} dIrrs should
be verified by observations at very long radio wavelengths to
possibly reveal an aged population of CR electrons radiating in weak
magnetic fields, or by Faraday rotation observations applied to
background sources. These possibilities are only now becoming possible with the
availability of the LOFAR and SKA pathfinders (Morganti et al.~\cite{morganti10}, 
Beck~\cite{beck10}), as well as the EVLA.

\subsection{Production of large-scale fields}
\label{s:dynamo}

We demonstrated that magnetic fields in the LG dIrrs (if
detectable) are not more ordered than in typical spiral
galaxies (Sect.~\ref{s:magnetic}) and must be dominated by
the random component. This conclusion suggests that LG dwarfs might not fulfil
the evolutionary scenario of Arshakian et al.~(\cite{arshakian09})
that local and moderately distant ($z\approx 1$) dwarfs can produce highly
coherent (unidirectional) fields.

The generation of coherent (regular) fields requires large-scale
dynamo process to operate, hence helical turbulence produced by
supernovae explosions and Coriolis forces (Widrow~\cite{widrow02}).
The dynamo efficiency is approximately described by the dynamo
number
$$D=9\frac{H^2 \Omega}{v^{2}_t} r \frac{\partial \Omega}{\partial r},$$
which depends on the vertical scale height of the galactic disk
$H$, turbulent velocity $v_t$, an angular rotation $\Omega$, and the
velocity shear ($r\partial \Omega /\partial r$). The large-scale
$\alpha$-$\Omega$ dynamo works only if $D$ exceeds a critical number
of about 9-11, depending on the details of the gas flow pattern. For the
LG dwarfs, we estimated the maximum rotational velocities $v_{rot}$ from
the galactic rotation curves or velocity dispersions (see
Table\,\ref{t:bfields} with notes). 
As some galaxies show evidence of gravitational interactions, these esimates 
are only approximate. For example, for strongly interacting NGC\,4449 
Valdez-Guti\'errez et al. (\cite{valdez02}) estimate that 
the systematic rotation on the receding side of the galaxy is about 
40\,km\,s$^{-1}$ at $2\arcmin$ radius (from H$\alpha$ measurements). 
On the same portion of the galaxy, the \ion{H}{i} observations indicate a 
rotation of about 30\,km\,s$^{-1}$ (Martin \cite{martin98}). 
The typical maximum rotational velocity for dwarfs is
about 30\,km\,s$^{-1}$ at a radius
$r=2$\,kpc. For differential rotation and flat rotation curve, this
corresponds to a shear of 15\,km\,s$^{-1}$\,kpc$^{-1}$. Taking
$v_t=10$\,km\,s$^{-1}$, $H=0.5$\,kpc (cf. Elstner et 
al.~\cite{elstner09}), we obtain a dynamo number of $D=5$, which is
subcritical and hence in agreement with our radio observations. If this 
rotation were not differential (e.g. more chaotic), then $D$ would be
even smaller. This is confirmed by direct simulations of a
supernova-driven dynamo by Gressel et al.~(\cite{gressel08}), who did
not find large-scale dynamo action within a galactic disk of 
corresponding shear 20\,km\,s$^{-1}$\,kpc$^{-1}$. Recent MHD
simulations of a cosmic-ray driven dynamo by Siejkowski et 
al.~(\cite{siejkowski10}) show that the production of regular fields in
dwarf galaxies requires mainly fast rotation. The velocity shear is
necessary but influences the dynamo efficiency much less. 

\begin{figure}[t]
    \begin{center}
        \includegraphics[width=7cm]{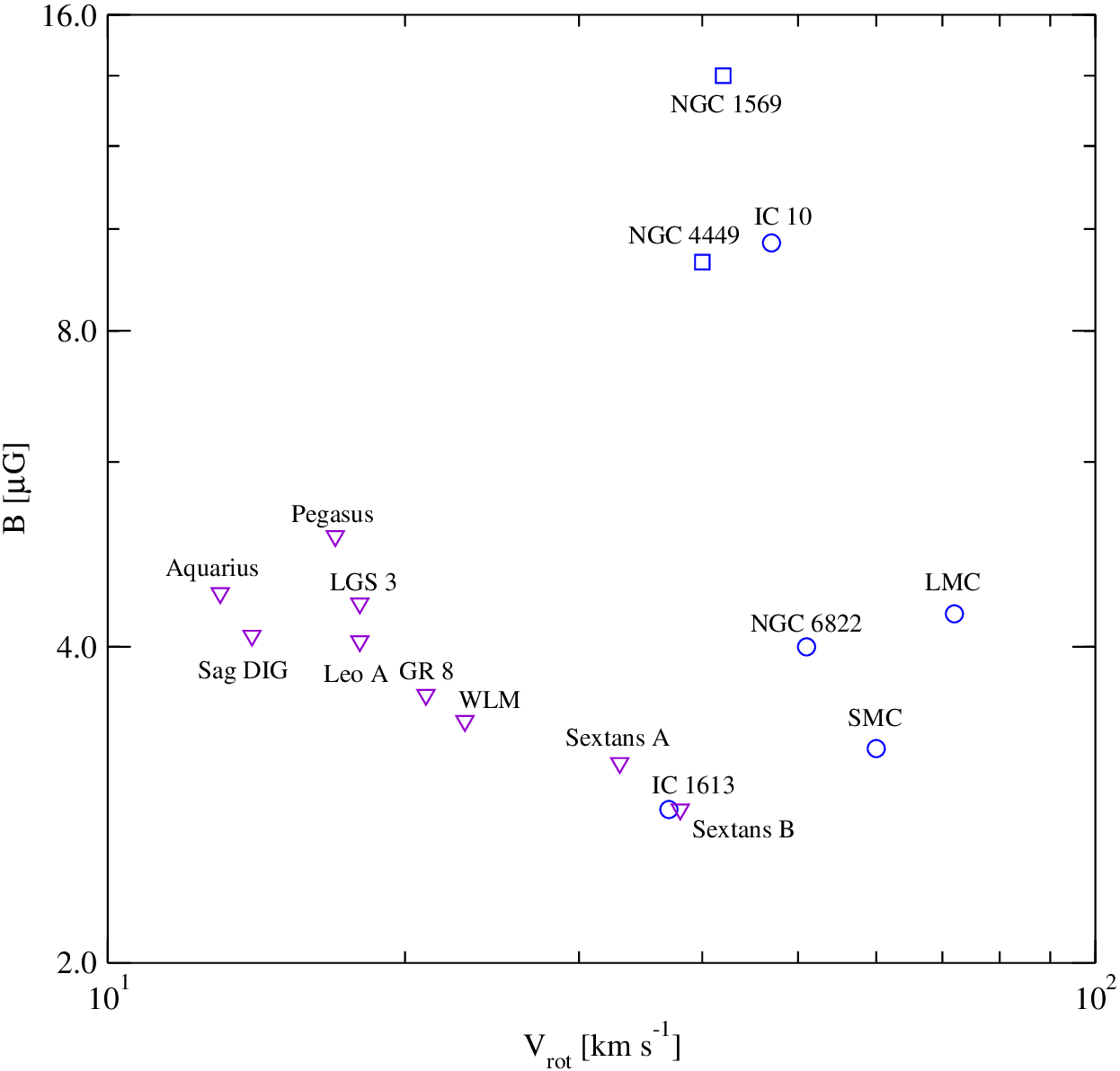}
    \end{center}
    \caption{
Total magnetic field strengths plotted against rotational velocities of dwarfs.
}
\label{f:vrot}
\end{figure}

In typical and even the most starbursting LG dwarf (IC\,10), the 
ordered magnetic fields have indeed been observed 
so far not to show any large-scale structure that could resemble MHD
dynamo fields (Chy\.zy et al.~\cite{chyzy03}). This confirms the
earlier suggestion by Chy\.zy et al.~(\cite{chyzy03}) that small
galaxies should mainly produce random magnetic fields maintained by
turbulent gas motions. The large irregular dwarf NGC\,4449 with its
dynamo-like field could be a special case
(Sect.\,\ref{s:evolution}). Rotation in dwarfs is usually not only
slower but also more chaotic than in spiral galaxies (e.g. van
Eymeren et al.~\cite{vaneymeren09}). Larger velocity dispersion and star
formation feedback may destroy the regular pattern of magnetic
fields and lower the field regularity. No MHD simulations have 
hitherto addressed these possibilities.

We do not observe a systematic dependence of the
maximum rotational velocity on the total magnetic field strength
in dwarfs (Fig.\,\ref{f:vrot}). For slow rotation
($v_{rot}<40$\,km\,s$^{-1}$), all dwarf galaxies show weak fields
($B<4\,\mu$G). Above a velocity of 40\,km\,s$^{-1}$, dwarfs have
either stronger or weaker fields that seems to mainly depend
on the actual value of the SFR. Indeed, NGC\,6822, LMC, and SMC
rotate at least as fast as the starbursting galaxies IC\,10,
NGC\,1569, and NGC\,4449, but they have less active and widespread
star formation.

According to a study of magnetic fields within the disk of the
large spiral galaxy NGC\,4254 (Chy\.zy~\cite{chyzy08}), the random
magnetic field scales with the far-infrared based
$\Sigma$SFR as a powerlaw with an exponent of $0.26\pm0.01$.
A similar relation with exponent of $0.25\pm0.04$ has also been found 
for LG dIrrs (Sect.\,\ref{s:sfr}). We are currently collecting
data on galaxies with properties in-between those of the dwarfs studied in
this paper and spiral galaxies to confirm these findings for
a larger sample of galaxies.

\subsection{Radio-infrared relation}
\label{s:radiofir}

\begin{figure}
    \begin{center}
        \resizebox{9cm}{!}{\includegraphics[angle=0]{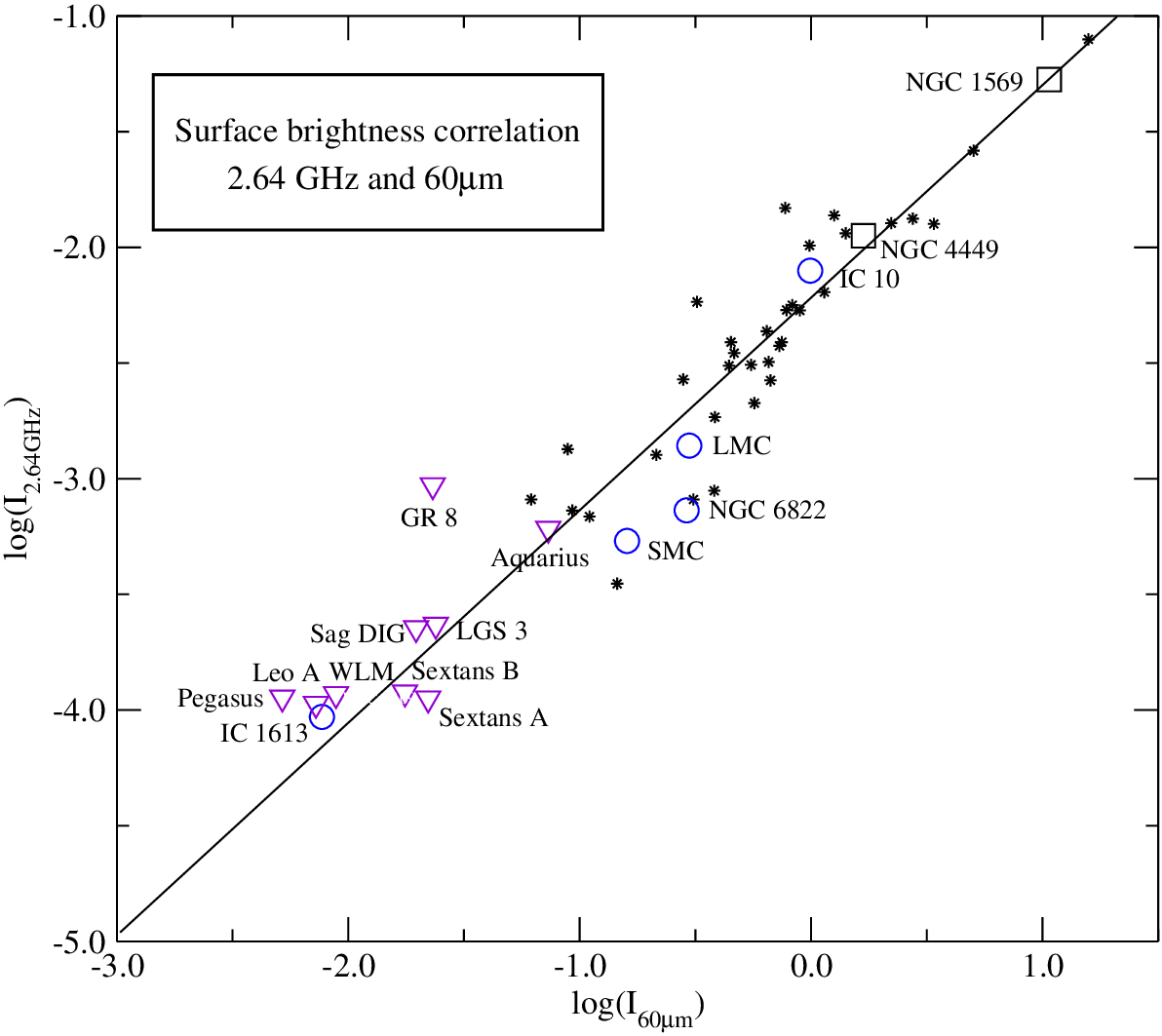}}
    \end{center}
    \caption{
The radio-infrared correlation diagram for LG dwarfs and comparison dwarfs
(plotted with the same symbols as in Fig.\,\ref{f:corel}), and for the sake
of reference, the sample of galaxies observed by de Jong~(\cite{deJong})
plotted as stars. The surface brightness at 2.64~GHz and at 60$\mu$m
is used (in Jy/$\sq\arcmin$). The solid line is an orthogonal fit to the
reference galaxies.
}
\label{f:radiofir}
\end{figure}

We constructed for the first time a radio-infrared correlation
diagram for dwarf galaxies (Fig.\,\ref{f:radiofir}), extending this
relation to the lowest limit of galactic mass of about several $10^7
\,M_{\sun}$, which is 10 times lower than previously studied 
(Chy\.zy et al.~\cite{chyzy07}). We used the surface brightness
radio emission at 2.64\,GHz (or its limits) and the infrared surface
brightness emission at $60\,\mu$m (Table~\ref{t:bfields}). The
relation for detected dwarfs is closely represented by the powerlaw
with a slope of $0.91\pm0.08$ and a correlation of $r=+0.91$
($P\ll1\%$) determined for a sample of bright galaxies observed by
de Jong~(\cite{deJong}) with the 100\,m NRAO telescope. Hence,
typically unevolved, low-mass stellar systems in our local
neighbourhood reveal similar physical conditions for star formation,
magnetic field, and cosmic-ray generation processes as the massive
spirals (e.g. We\.zgowiec et al.~\cite{wezgowiec07}). A 
similar relation was extended to young, high-redshift galaxies 
(Seymour et al.~\cite{seymour08}).

However, it cannot be excluded that some radio deficiency might be present in
some dwarfs of weak infrared radiation, as only
the upper limits of their radio brightness are plotted in
Fig.\,\ref{f:radiofir}. The loss of magnetic fields and CRs by
galactic outflows in dwarfs and subsequently their weaker
synchrotron emission could be counterbalanced by the lower content of
dust as suggested for blue compact dwarfs by Klein at al.~(\cite{klein91}). 
In this case, the radio-infrared relation could be
preserved. Because dIrrs at the low-end of the radio flux also have
lower metallicity (which scales with the object's mass) the shortage
of dust is indeed likely to occur.

According to the radio-infrared relation, the radio-undetected
dwarfs from our LG sample are expected to have a radio brightness at
2.64\,GHz of about 1\,mJy per $1\arcmin$ beam or less. This requires a few times 
higher detecting sensitivity than we have achieved with the Effelsberg
telescope. However, these fluxes are below the confusion limit of a
single-dish antenna and demand observations of higher resolution.
Objects with such weak and extended radio emission are extremely
difficult to detect at higher frequencies. Studies at lower frequencies would
thus require a high-sensitivity radio interferometer such as LOFAR. If
these dwarf galaxies hosted primarily regular magnetic fields of about
$1\,\mu$G strength (Arshakian et al.~\cite{arshakian09}), then they
could also be detected by the background Faraday rotation method
with the upcoming LOFAR or SKA interferometers.

\section{Summary and conclusions}
\label{s:summary}

We have performed a sensitive search for radio emission in a
statistically unbiased sample of the Local Group irregular and dwarf
irregular galaxies available at the site of 100-m Effelsberg
telescope at 2.64\,GHz. We have compared these 12 galaxies to other LG
objects, i.e., LMC and SMC, and the starburst dwarfs NGC\,4449 and
NGC\,1569, for which magnetic fields have already been observed.
Higher frequency (4.85\,GHz) observations were used to search for polarized
emission in the five most luminous dwarfs in the infrared domain.

We found the following:

\begin{itemize}

\item The LG dIrrs closely represent the local volume of the Universe 
and display star-formation characteristics similar to those of other nearby groups 
of dwarfs (Sect.\,\ref{s:evolution}). Only 25\% of the LG dwarfs (3 out of 12)
were detected at 2.64\,GHz. We argued that the radio-undetected dwarfs must 
be intrinsically radio weak.

\item The total magnetic fields in LG dwarfs are very weak: the mean value of
the equipartition magnetic field strength of all objects is $<4.2\pm
1.8\,\mu$G, taking into account the upper limits for
radio-undetected dwarfs (9 out of 12). This value is almost three
times weaker than the estimated mean for typical spiral galaxies.
The strongest total field of $10\,\mu$G is observed in the starburst
dwarf IC\,10. We found that the magnetic field strength does not
correlate with the dwarf rotational velocity.

\item Among the three radio-detected dwarfs, the strength of the ordered field component
is in the range $0.4-0.9\,\mu$G and the ordered-to-random field 
component ratio is about 0.2. These values are smaller than in typical 
spiral galaxies indicating that the production of magnetic fields in 
dwarfs is probably not maintained by the large-scale dynamo process.

\item The production of the total magnetic field in dwarf systems appears 
to be controlled mainly by the star-formation surface density ($\vec{B}\propto
\Sigma$SFR$^{0.30\pm0.04}$), or the gas density ($\vec{B}\propto
\Sigma \rho^{0.47\pm0.09}$), as for spiral galaxies
(Chy\.zy~\cite{chyzy08}, Krause~\cite{krause09}). We note a somewhat
steeper Schmidt law (a slope $N=1.8$) for our LG dwarfs than in
typical large-mass disks. Among all the dwarfs, we also found
systematically stronger magnetic fields in objects of higher {\em
global} SFR ($\vec{B}\propto\,$SFR$^{0.25\pm0.06}$).

\item Stronger disk-averaged magnetic fields ($>4\,\mu$G) were observed in
dIrrs of extreme characteristics only (e.g. NGC\,4449, NGC\,1569,
and the LG dwarf IC\,10). They are more evolved objects of generally much higher
metallicity and global SFR than the majority of LG dwarf population. They also usually
show clear signs of current or recent gravitational interactions.

\item We propose that a coeval magnetization of the IGM around primordial
galaxies occurs with a metal enrichment caused by galactic outflows maintained by
stellar winds and supernovae. However, our modelling of the stellar population 
synthesis (code Starburst99) and expansion of a blowing bubble indicate
that not dIrrs but more massive galaxies of LBG properties
can efficiently magnetize the IGM at high ($z>3$)
redshift. If the current understanding and modelling of metal
enrichment is valid, we expect that the most efficient seeding of the IGM is that produced by
LBG galaxies with magnetic field strengths of about $0.9-0.6$\,nG at up
to distances of 160--530\,kpc at redshifts 5--3, respectively. We
show that several times weaker fields and shorter distances are
expected from primordial dwarf galaxies. 

\item The sizes of blowing bubbles around primordial galaxies might be 
slightly enlarged (by up to 26\%) if in addition to the thermal pressure 
the pressure of CRs is included.

\item We also predict that around local, most star-forming dwarf galaxies,
the surrounding IGM is magnetized up to about 0.1\,$\mu$G to within a distance of 
about 5\,kpc. These predictions should be verified by LOFAR, EVLA, and SKA
observations.

\item The dIrrs of different SFR follow the far-infrared relationship determined
for high surface brightness spiral galaxies (with a slope $0.91\pm0.08$),
showing that similar processes regulate star formation, synchrotron
emission, and production of magnetic fields in low-mass dwarfs and large
spirals. From the radio-infrared relation, we
predict for the radio-undetected dwarfs in our sample a mean total field
strength as small as about $1\,\mu$G .

\end{itemize}

\begin{acknowledgements}
We are grateful to Prof. M. Urbanik for permission to present
4.85\,GHz data of IC\,1613 from our common observation in 2002, and
to Dr M. Soida for valuable discussions. K.T. Chy\.zy is grateful to
Prof. Claus Leitherer for his help in the compilation and setup of
the Starburst99 code. We thank Dr. M. Krause for careful reading of
the manuscript and an anonymous referee for helpful comments and 
suggestions. This work was supported by the Polish Ministry of
Science and Higher Education, grant 2693/H03/2006/31,
3033/B/H03/2008/35, and from research funding from the European
Community's sixth Framework Programme under RadioNet R113CT 2003
5058187. We acknowledge the use of the HyperLeda
(http://leda.univ-lyon1.fr) and NED (http://nedwww.ipac.caltech.edu)
databases.
\end{acknowledgements}

\appendix
\section{Derivation of magnetic field strengths}

According to the theory of synchrotron emission (e.g. Pacholczyk 
\cite{pacholczyk70}), the total synchrotron intensity and the degree 
of linear polarization obtained from our radio polarimetric 
observations can be used to calculate the strengths of the total 
($B_{tot}$) and regular ($B_{reg}$) magnetic field. With the assumption of 
equipartition between the energy densities of the magnetic field and cosmic rays 
($\varepsilon_{CR} = \varepsilon_{B_{tot}} = {B_{tot}}/{8\pi}$), 
the total magnetic field is (Beck \& Krause~\cite{beckra})
\begin{equation}
B_{tot} = \left[\frac{4\pi(2\alpha_n+1)(K_0+1)I_{n}E_p^{1-2\alpha_n}(\frac{\nu}{2c_1})^{\alpha_n}}{(2\alpha_n-1)c_2\alpha_nLc_3}\right]^{1\over{\alpha_n+3}},
\end{equation}
where $K_0$ is the constant ratio of proton to electron number densities, $I_n$ is the 
nonthermal intensity, and $\alpha_n$ is the 
mean synchrotron spectral index, $L$ denotes the pathlength through the synchrotron emitting medium, 
$E_p$ is the proton rest energy, and $c_1$ is a constant defined as
\begin{equation}
c_1 = \frac{3e}{4\pi m_e^3c^5} = 6.2648 \times 10^{18} {\rm erg}^{-2}{\rm s}^{-1}{\rm G}^{-1}.
\end{equation}
The constants $c_2$ and $c_3$ depend on the spectral index and the inclination of the magnetic field, 
respectively
\begin{equation}
c_2(\alpha_n) = {1\over4}c_3 \frac{(\alpha_n+{5\over3})}{(\alpha_n+1)} 
\Gamma\left[{(3\alpha_n+1)\over6}\right]\times\Gamma\left[{(3\alpha_n+5)\over6}\right],
\end{equation}
where $c_3 = (\cos i)^{\alpha_n+1}$. This is true for a region where the field is totally regular and has 
a constant inclination $i$ with respect to the sky plane. If the synchrotron intensity is averaged over a 
large volume (as in estimating the mean field strength), the value of $c_3$ has to be replaced by its 
average over all occurring values of $i$. For a totally turbulent field, one needs to use 
$c_3 = (2/3)^{\alpha_n+1)/2}$.

\begin{table}[t]
\caption{The dependence of the total magnetic field (in $\mu$G) of NGC\,6822 on the 
synchrotron pathlength ($L$) and the proton-to-electron ratio ($K_0$). The original value 
is marked in boldface.}
\begin{center}
\begin{tabular}{ccccccc}
\hline
\hline
$K_o/L$ [kpc]	&	1	&	1.5	&	2	&	2.5	&	3	\\
\hline
50		&	3.97	&	3.57	&	3.34	&	3.14	&	2.99	\\	
75		&	4.39	&	3.96	&	3.70	&	3.47	&	3.31	\\
100		&	4.73	&	4.26	&   {\bf 3.98}	&	3.74	&	3.57	\\
125		&	5.00	&	4.51	&	4.21	&	3.95	&	3.77	\\
150		&	5.24	&	4.72	&	4.41	&	4.14	&	3.95	\\
\hline
\end{tabular}
\end{center}
\label{n6822fields}
\end{table}

To estimate the strength of the regular magnetic field in the sky plane, we can use the observed 
nonthermal degree of polarization (Segalovitz et al.~\cite{segalovitz76})
\begin{equation}
P_{nth} = \left(\frac{3\gamma+3}{3\gamma+7}\right)\times \left[1+\frac{(1-q)\pi^{1\over2}\Gamma[(\gamma+5)/4]}{2q\Gamma[(\gamma+7)/4]F(i)}\right]^{-1},
\end{equation}
where
\begin{equation}
F(i) = {1\over2\pi}\int_0^{2\pi}(1-{\rm sin}^2i\,{\rm sin}^2\theta)^{(\gamma+1)/4}{\rm d}\theta.
\end{equation}
In the above formulae, $q^2/(1+\gamma)=B_{reg}/B_{turb}$, $\gamma = 2\alpha_n+1$, and $\theta$ is 
the azimuthal angle.

In calculating magnetic field strengths, some uncertainties can be introduced by assuming a  
proton-to-electron number density ratio, as well as a geometry for the galaxy.
We assume ellipsoidal geometries of the studied galaxies,
using their minor axes as the synchrotron pathlength.
An ellipsoidal approximation of the geometry of a dwarf galaxy was 
applied by Chy\.zy et al.~(\cite{chyzy03}) to derive the magnetic
field strength for NGC\,6822 from the 4.85\,GHz data.
To test the influence of these parameters, we performed calculations of the 
magnetic field in NGC\,6822, varying $K_o$ in the range of 50-150 and $L$ in 
the range of 1-3\,kpc. The results are presented in Table~\ref{n6822fields}. 
The largest deviation in our calculations was 32\% (see Table~\ref{n6822fields}), 
while typically it did not exceed 15\%. Therefore, varying these parameters by 50\%, 
as we also did for the nonthermal flux and the degree of polarization, to calculate 
the global error in the magnetic field strengths (see Sect.~\ref{s:magnetic}), provides 
us with errors that thoroughly estimate the possible uncertainties.
We also tested whether the magnetic field strength depends on galaxy distance, thus on the 
location within the Local Group with respect to the Milky Way. For all LG dIrrs, the
correlation coefficient $r=-0.37$ with $P=23\%$, which means that the distance has 
no statistically significant influence.

\end{document}